\documentclass[preprint,aps,prd,groupedaddress,showpacs,nofootinbib]{revtex4}
 
\usepackage{amsmath}
\usepackage{graphicx}
\usepackage{epsfig}
\usepackage{bm}

\allowdisplaybreaks

\def\lQ{\Lambda_{\rm QCD}}
\def\als{\alpha_{\rm s}}
\def\alVs{\alpha_{V_s}}
\def\alVo{\alpha_{V_o}}
\def\siml{{\ \lower-1.2pt\vbox{\hbox{\rlap{$<$}\lower6pt\vbox{\hbox{$\sim$}}}}\ }}
\def\simg{{\ \lower-1.2pt\vbox{\hbox{\rlap{$>$}\lower6pt\vbox{\hbox{$\sim$}}}}\ }}

\newcommand{\be}{\begin{equation}}
\newcommand{\ee}{\end{equation}}
\newcommand{\bea}{\begin{eqnarray}}
\newcommand{\eea}{\end{eqnarray}}
\newcommand{\nn}{\nonumber}

\begin{document}

\preprint{\tt BNL-NT-08/7, IFUM-916-FT}

\title{Static quark-antiquark pairs at finite temperature}
\author {Nora Brambilla, Jacopo Ghiglieri, Antonio Vairo}
\affiliation{Dipartimento di Fisica dell'Universit\`a di Milano and INFN\\
via Celoria 16, 20133 Milan, Italy}
\author {P\'eter Petreczky}
\affiliation{RIKEN-BNL Research Center \& Physics Department \\ 
Brookhaven National Laboratory, Upton, NY 11973, USA}

\date{\today}

\begin{abstract}
In a framework that makes close contact with modern effective field theories 
for non-relativistic bound states at zero temperature, we study the real-time 
evolution of a static quark-antiquark pair in a medium of gluons and light 
quarks at finite temperature. For temperatures ranging from  values larger 
to smaller than the inverse distance of the quark and antiquark, $1/r$, and 
at short distances, we derive the potential between the two static sources, 
and calculate their energy and thermal decay width.  Two mechanisms contribute 
to the thermal decay width: the imaginary part of the gluon self energy 
induced by the Landau damping phenomenon, and the quark-antiquark color singlet to color 
octet thermal break up. Parametrically, the first mechanism dominates for temperatures 
such that the Debye mass is larger than the binding energy, while the latter, 
which we quantify here for the first time, dominates for temperatures such that 
the Debye mass is smaller than the binding energy. 
If the Debye mass is of the same order as $1/r$, our results are in agreement 
with a recent calculation  of the static Wilson loop at finite temperature. 
For temperatures smaller than $1/r$, we find new contributions to the potential, 
both real and imaginary, which may be relevant to understand the onset of 
heavy quarkonium dissociation in a thermal medium.
\end{abstract}


\maketitle

\section{Introduction}
The study of heavy quark-antiquark pairs in a thermal medium at temperature $T$ 
has received a lot of attention since it was suggested that quarkonium dissociation due to color screening may be 
a striking signature of the quark-gluon plasma formation \cite{Matsui:1986dk}.
Based on this idea and the assumption that medium effects can be understood in terms of a temperature dependent
potential, the problem of quarkonium dissociation has been addressed in terms of potential models with
screened temperature-dependent potentials over the past 20 years (see e.g. 
Refs.~\cite{Karsch:1987pv,Digal:2001iu,Mocsy:2005qw} for some representative works). 
A derivation from QCD of the in-medium quarkonium potential has not appeared  
in the literature so far and expectedly not all medium effects can be incorporated into a potential. 
A first step towards a QCD derivation of the quarkonium potential at finite temperature  
has been a recent calculation \cite{Laine:2006ns} of the static Wilson loop 
in the imaginary-time formalism at order $\als$. After analytical continuation to real time, 
the calculation shows a real part, which is a screened Coulomb potential, and an imaginary part 
that may be traced back to the scattering of particles in the medium 
carrying momenta of order $T$ with space-like gluons, a phenomenon also known as Landau damping. 
Some applications can be found in \cite{Laine:2007gj,Laine:2007qy}. 
First principle calculations of quarkonium properties at finite temperature 
include calculations of Euclidean correlation functions 
in lattice QCD and the reconstruction of the corresponding spectral functions  
using the Maximum Entropy Method. At present, however, a reliable determination of the
quarkonium spectral functions from the lattice data appears very difficult 
due to statistical errors and lattice discretization effects (see discussion in Ref.~\cite{Jakovac:2006sf} 
and references therein). 

In this work, we will study static quark-antiquark pairs in a thermal bath 
in real-time formalism (see e.g. \cite{LeBellac:1996}) and in a framework that makes close contact with 
effective field theories (EFTs) for non-relativistic bound states at $T=0$ \cite{Brambilla:2004jw}. 
In this framework, we will address the problem of defining and deriving the potential 
between the two static sources, and we will calculate their energy and 
thermal decay width. We will describe the system for temperatures that range from larger 
to smaller values with respect to the inverse distance of the quark and antiquark, $1/r$. 
In some range, we will agree with previous findings; for temperatures lower 
than the inverse distance of the quark and antiquark, we will find new contributions
to the potential, both real and imaginary, with a non trivial analytical structure.
In particular, we will point out the existence of a new type of process that 
contributes to the quark-antiquark thermal decay width besides the Landau damping.

We will deal with static quarks only. The static case is relevant also for the 
study of bound states made of quarks with a large but finite mass $m$, like quarkonia, 
in a thermal medium. Quarkonium is expected to exist in the medium if  
the temperature and the other thermodynamical scales are much lower than $m$.
In this situation, one may consistently integrate out the mass from QCD and expand 
order by order in $1/m$. The leading order of the expansion corresponds 
to QCD with a static quark and a static antiquark. Higher-order corrections in $1/m$ 
may be systematically included in the framework of non-relativistic QCD (NRQCD)  
\cite{Caswell:1985ui}.

Bound states at finite temperature are systems characterized by many energy scales.
There are the thermodynamical scales that describe the motion of the particles in the thermal bath:
the temperature scale $T$ (we will not distinguish between $T$ 
and multiple of $\pi T$), the Debye mass $m_D$, which is the scale of the screening 
of the chromoelectric interactions, and lower energy scales. 
In the weak-coupling regime, which we will assume throughout this work, 
one has $m_D \sim gT \ll T$. Moreover, there are the scales typical of the 
bound state. In the case of a system of two static sources, the scales  may be identified 
with the inverse of the quark-antiquark distance $r$ and the static potential, the first being 
much larger than the second. We will also assume that $1/r$ is much larger 
than the typical hadronic scale $\lQ$, i.e. we will concentrate on the short-distance 
part of the potential. This may be the part of the potential relevant for the lowest quarkonium 
resonances like the $J/\psi$ or the $\Upsilon(1S)$, which are the most tightly bound states.
Thermodynamical and bound-state scales get entangled and different hierarchies are 
possible. Bound states are expected to dissolve in the bath at temperatures such that $m_D$ is 
larger than the typical inverse size of the bound state.
Hence we will concentrate on the situation $1/r \simg m_D$
(i.e. $1/r \gg m_D$ or $1/r \sim m_D$), and distinguish
between the two cases $1/r \gg T$ and $1/r \ll T$.

The paper is organized as follows. Sections ~\ref{secQCD} and \ref{secHTL} are introductory:
they deal with QCD with static sources, which we call static QCD for short, but do not include bound states. 
In Sec.~\ref{secQCD}, we write the quark and gluon propagators in static QCD at finite $T$. 
In Sec. \ref{secHTL}, we summarize one-loop finite $T$ contributions in static QCD  
that are relevant to the present work. 
In Sec.~\ref{secWpNRQCD}, we introduce the relevant EFTs and calculate the static potential 
in a situation where the inverse distance  between the static quark and antiquark 
is larger than the temperature of the thermal bath: $1/r \gg T \gg m_D$. When $T$ is as small 
as the binding energy, we also calculate the leading thermal contribution to the static energy and the decay width. 
In Sec.~\ref{secpQCD}, we provide an alternative derivation of the potential in perturbative QCD. 
In Sec.~\ref{secSpNRQCD}, we calculate the static potential in the situation  $T \gg 1/r \simg m_D$. 
In Sec.~\ref{secCon}, we summarize and discuss our results and list some possible developments.

\section{Static QCD at finite $T$}
\label{secQCD}
We consider here QCD with a static quark and antiquark; in particular, we write the quark, 
antiquark  and the gluon propagators at finite $T$. To simplify the notation, we will drop 
the color indices from the propagators. Throughout the paper, the complex time contour 
for the evaluation of the real-time thermal expectation values goes from a real initial time $t_{\rm i}$ 
to a real final time $t_{\rm f}$, from $t_{\rm f}$ to $t_{\rm f}-i0^+$, from 
$t_{\rm f}-i0^+$ to $t_{\rm i}-i0^+$ and from $t_{\rm i}-i0^+$ to $t_{\rm i}-i/T$.
The propagators will be given with this conventional choice of contour. 
Furthermore, the following notations will be used. We indicate thermal averages as 
\be
\langle O  \rangle_T = \frac{ {\rm Tr} \{ e^{-H/T}\, O  \}}
{ {\rm Tr} \{ e^{-H/T} \} }, 
\ee
where $H$ is the Hamiltonian of the system.  We also define
\bea
&&n_{\rm F}(k^0) = \frac{1}{e^{k^0/T}+1},
\\
&&n_{\rm B}(k^0) = \frac{1}{e^{k^0/T}-1}.
\eea

\subsection{Quark propagator}
In order to show the behaviour of static sources in a thermal bath, 
it may be useful to consider first a quark (or antiquark) with a large 
but finite mass $m$, $m\gg T$, and then perform the $m \to \infty$ limit.

We define the propagators 
\bea 
&& S_{\alpha\beta}^{>}(x) = \langle \psi_\alpha(x)\psi^\dagger_\beta(0)\rangle_T,
\\
&& S_{\alpha\beta}^{<}(x) = - \langle \psi^\dagger_\beta(0) \psi_\alpha(x)\rangle_T,
\eea
where $\psi$ is the Pauli spinor field that annihilates the fermion (in the following, 
the Pauli spinor field that creates the antifermion will be denoted $\chi$).
The free propagators, 
\be
S_{\alpha\beta}^{>\,(0)} = \delta_{\alpha\beta}S^{>\,(0)}, \qquad 
S_{\alpha\beta}^{<\,(0)} = \delta_{\alpha\beta}S^{<\,(0)},
\ee 
satisfy the equations (in momentum space:  $\displaystyle S(k) = \int d^4x \, e^{ikx} \, S(x)$): 
\bea 
k^0 S^{>\,(0)}(k) = m S^{>\,(0)}(k), 
\label{eqmot}
\\
k^0 S^{<\,(0)}(k) = m S^{<\,(0)}(k),
\label{eqmot1}
\eea
where we have neglected corrections of order $1/m$ or smaller: they will eventually vanish 
in the $m\to\infty$ limit.

If the heavy quarks are part of the thermal bath, they satisfy the Kubo--Martin--Schwinger relation:
\be
S^{<\,(0)}(k) = - e^{-k^0/T}  S^{>\,(0)}(k).
\label{KMSfermion}
\ee
From the equal-time canonical commutation relation it follows the sum rule
\be
\int \frac{dk^0}{2\pi}\left(S^{>\,(0)}(k) -S^{<\,(0)}(k)   \right) =1.
\label{sumrulefermion}
\ee
The solutions of the equations (\ref{eqmot})-(\ref{sumrulefermion}) are 
\bea
S^{>\,(0)}(k) = (1-n_{\rm F}(k^0))\, 2\pi\delta(k^0-m),
\label{Spm}
\\
S^{<\,(0)}(k) =  -n_{\rm F}(k^0)\, 2\pi\delta(k^0-m).
\label{Smm}
\eea
The spectral density $\rho^{(0)}_{\rm F}$ is given by
\be 
\rho^{(0)}_{\rm F}(k) = S^{>\,(0)}(k) - S^{<\,(0)}(k) = 2\pi\delta(k^0-m),
\ee
and the free propagator, 
\be 
 S^{(0)}(x) = \theta(x^0) S^{>\,(0)}(x) -  \theta(-x^0) S^{<\,(0)}(x), 
\ee
is given in momentum space by
\be 
 S^{(0)}(k) = \frac{i}{k^0-m+i\epsilon} -  n_{\rm F}(k^0) \, 2\pi \delta(k^0-m).
\ee

In the static limit $m \to \infty$, the propagators simplify because 
$n_{\rm F}(m) \to 0$ for $m \to \infty$. Moreover, we may 
get rid of the explicit mass dependence by means of the field redefinition
$\displaystyle \psi \to \psi e^{-imt}$, which 
amounts to change $k^0-m$ to $k^0$ in the expressions 
for the propagators and the spectral density; they read now 
\bea
S^{>\,(0)}(k) = 2\pi\delta(k^0),
\label{Sp}
\\
S^{<\,(0)}(k) = 0,
\label{Sm}
\\
S^{(0)}(k) = \frac{i}{k^0+i\epsilon},
\label{SS}
\\
\rho^{(0)}_{\rm F}(k) = 2\pi\delta(k^0).
\label{spectralS}
\eea
The free static propagator is the same as at zero temperature.
On the other hand, if we would have assumed from the beginning that  
$S^{<\,(0)}(k) = 0$, i.e. that there is no backward propagation 
of a static quark (in agreement with  the Kubo--Martin--Schwinger formula 
in the $m\to\infty$ limit) then, together with the equations  of motion  $k^0 S^{>\,(0)}(k) = 0$, 
$k^0 S^{<\,(0)}(k) = 0$ (obtained after removing $m$ via field redefinitions) and  
the sum rule (\ref{sumrulefermion}), we would have obtained Eqs. (\ref{Sp}), (\ref{SS})
and (\ref{spectralS}). 

The real-time free static propagator for the quark reads 
\be
{\bf S}_{\alpha\beta}^{(0)}(k) = 
\delta_{\alpha\beta}
\left(
\begin{matrix}
&&\hspace{-2mm}S^{(0)}(k)  
&&\hspace{-2mm}S^{<\,(0)}(k) \\ 
&&\hspace{-2mm}S^{>\,(0)}(k)  
&&(S^{(0)}(k))^*
\end{matrix}
\right)
= 
\delta_{\alpha\beta}
\left(
\begin{matrix}
&&\hspace{-2mm}\displaystyle\frac{i}{k^0+i\epsilon}  
&&0 \\ 
&&2\pi\delta(k^0)  
&&\displaystyle\frac{-i}{k^0-i\epsilon}
\end{matrix}
\right),
\label{Squarkstatic}
\ee
and for the antiquark
\be
{\bf S}_{\alpha\beta}^{(0)}(k) 
= 
\delta_{\alpha\beta}
\left(
\begin{matrix}
&&\hspace{-2mm}\displaystyle\frac{i}{-k^0+i\epsilon}  
&&0 \\ 
&&\hspace{-2mm}2\pi\delta(k^0)  
&&\displaystyle\frac{-i}{-k^0-i\epsilon}
\end{matrix}
\right).
\label{Santiquarkstatic}
\ee
The main observation here is that, since the $[{\bf S}_{\alpha\beta}^{(0)}(k)]_{12}$
component vanishes, the static  quark (antiquark) fields labeled ``2'' 
never enter in any physical amplitude, i.e. any amplitude that has the 
physical fields, labeled ``1'', as initial and final states. 
Hence, when considering physical amplitudes, the static fields ``2'' 
decouple and may be ignored.

The propagator ${\bf S}_{\alpha\beta}^{(0)}$ may be written in a diagonal form as 
\be
{\bf S}_{\alpha\beta}^{(0)}(k) = {\bf U}^{(0)} 
\left(
\begin{matrix}
&&\hspace{-2mm}\displaystyle\frac{i}{k^0+i\epsilon}  
&&0 \\ 
&&0  
&&\displaystyle\frac{-i}{k^0-i\epsilon}
\end{matrix}
\right)
{\bf U}^{(0)}\,,
\label{diagS0}
\ee
where
\be
{\bf U}^{(0)} = 
\left(
\begin{matrix}
&&\hspace{-2mm}1
&&0 \\ 
&&\hspace{-2mm}1  
&&~1
\end{matrix}
\right)\,,
\quad \hbox{and for further use}\quad 
[{\bf U}^{(0)}]^{-1} = 
\left(
\begin{matrix}
&&\hspace{-2mm}1
&&0 \\ 
&&\hspace{-2mm}-1  
&&~1
\end{matrix}
\right)\,.
\label{matrixU0}
\ee
Throughout the paper, we will use bold-face letters to indicate $2\times 2$ matrices 
in the real-time formalism.

\subsection{Gluon propagator}
\label{secgluonpropagator}
The gluon propagator in the real-time formalism can be written as \cite{LeBellac:1996}
\be
{\bf D}_{\alpha\beta}(k) = 
\left(
\begin{matrix}
&&\hspace{-2mm} \displaystyle D_{\alpha\beta}(k)
&&D^<_{\alpha\beta}(k)
\\
&&\hspace{-2mm} D^>_{\alpha\beta}(k) 
&&\displaystyle (D_{\alpha\beta}(k))^*
\end{matrix}
\right),
\label{realtimeD}
\ee
where
\bea
&& D_{\alpha\beta}^{>}(k) = \int d^4x \, e^{ik\cdot x} \, \langle A_\alpha(x)A_\beta(0)\rangle_T,
\label{Dp}
\\
&& D_{\alpha\beta}^{<}(k) = \int d^4x \, e^{ik\cdot x} \,\langle A_\beta(0) A_\alpha(x)\rangle_T,
\label{Dm}
\\
&& D_{\alpha\beta}(k) = \int d^4x \, e^{ik\cdot x} \,\left[
\theta(x_0)\langle A_\alpha(x)A_\beta(0)\rangle_T + \theta(-x_0)\langle A_\beta(0) A_\alpha(x)\rangle_T
\right].
\label{DD}
\eea
Gluons being bosonic fields, the Kubo--Martin--Schwinger relation reads
\be
D^<(k) = e^{-k^0/T}  D^>(k),
\label{KMSboson}
\ee
from which it follows that 
\bea 
&& D_{\alpha\beta}^{>}(k) = \left( 1 + n_{\rm B}(k^0) \right) \rho_{{\rm B}\,\alpha\beta}(k), 
\label{Dgn}
\\
&&D^<_{\alpha\beta}(k) = n_{\rm B}(k^0) \rho_{{\rm B}\,\alpha\beta}(k) ,
\label{Dln}
\eea
where 
\be 
\rho_{{\rm B}\,\alpha\beta}(k) = D_{\alpha\beta}^{>}(k) - D_{\alpha\beta}^{<}(k),
\ee
is the spectral density.

We may express ${\bf D}_{\alpha\beta}$ also in terms of the retarded and advanced propagators 
$D_{\alpha\beta}^{\rm R}$ and $D_{\alpha\beta}^{\rm A}$:
\bea
&& D_{\alpha\beta}^{\rm R}(k) = \int d^4x \, e^{ik\cdot x} \, 
\theta(x_0)\langle [A_\alpha(x),A_\beta(0)]\rangle_T,
\label{DR}
\\
&& D_{\alpha\beta}^{\rm A}(k) = - \int d^4x \, e^{ik\cdot x} \,
\theta(-x_0) \langle [A_\alpha(x), A_\beta(0)]\rangle_T;
\label{DA}
\eea
we have
\bea
&& \rho_{{\rm B}\,\alpha\beta}(k) = D_{\alpha\beta}^{\rm R}(k) - D_{\alpha\beta}^{\rm A}(k),
\label{rhoRA}
\\
&& D_{\alpha\beta}(k) =  D^{\rm R}_{\alpha\beta}(k) + D^{<}_{\alpha\beta}(k)
= D^{\rm A}_{\alpha\beta}(k) + D^{>}_{\alpha\beta}(k) 
\nn\\
&& \hspace{1.4cm}
= \frac{ D_{\alpha\beta}^{\rm R}(k) + D_{\alpha\beta}^{\rm A}(k)}{2}
+ \left(\frac{1}{2} + n_{\rm B}(k^0)\right) \rho_{{\rm B}\,\alpha\beta}(k). 
\label{DRA}
\eea

In the free case, in Coulomb gauge, the longitudinal and transverse propagators have the following 
expressions \cite{Landshoff:1992ne}:
\bea
{\bf D}^{(0)}_{00}(\vec k) &=& 
\left(
\begin{matrix}
&&\hspace{-2mm} \displaystyle \frac{i}{\vec{k}^2}
&&0
\\
&&\hspace{-2mm} 0
&&\displaystyle -\frac{i}{\vec{k}^2}
\end{matrix}
\right), 
\label{D000}
\\
{\bf D}^{(0)}_{ij}(k) &=& 
\left(\delta_{ij} - \frac{k^ik^j}{\vec k^2}\right)
\!\left\{\!
\left(
\begin{matrix}
&&\hspace{-2mm} \displaystyle \frac{i}{k^2 + i\epsilon}
&&\theta(-k^0) \, 2\pi\delta(k^2)
\\
&&\hspace{-2mm} \theta(k^0) \, 2\pi\delta(k^2)
&&\displaystyle -\frac{i}{k^2 - i\epsilon}
\end{matrix}
\right)
+ 2\pi\delta(k^2)\, n_{\rm B}(|k^0|)\,
\left(
\begin{matrix}
&&\hspace{-2mm} 1
&& 1
\\
&&\hspace{-2mm} 1
&& ~1
\end{matrix}
\right)\!
\right\}\!,
\nn\\
\label{D0ij}
\eea
where $k^2 = (k^0)^2- {\vec k}^2$.
Note that the longitudinal part of the gluon propagator in Coulomb gauge does not depend on the 
temperature. The temperature enters only the transverse part, which splits in the sum of a $T=0$ 
piece and a thermal one.

\subsection{Lagrangian}
The Lagrangian of QCD with a static quark, a static antiquark and $n_f$ massless quark fields $q_i$ is 
\be
{\cal L}  = 
- \frac{1}{4} F^a_{\mu \nu} F^{a\,\mu \nu} 
+ \sum_{i=1}^{n_f}\bar{q}_i\,iD\!\!\!\!/\,q_i 
+ \psi^\dagger i D_0 \psi  + \chi^\dagger i D_0 \chi,
\label{NRQCDstatic}
\ee
where $i D^0=i\partial_0 -gA^0$, $i{\vec{D}}=i\vec{\nabla}+g{\vec{A}}$, and
$i g F_{\mu \nu} = [D_\mu,D_\nu]$. The free static quark propagator is given 
by Eq. (\ref{Squarkstatic}), the free static antiquark propagator by Eq. (\ref{Santiquarkstatic})
and the free gluon propagator (in Coulomb gauge) by Eqs. (\ref{D000}) and (\ref{D0ij}).
Note that transverse gluons do not couple directly to static quarks.

\section{One-loop finite $T$ contributions in static QCD}
\label{secHTL}
Throughout this work, we will assume that $T$, $gT$ $\gg \lQ$; this enables 
us to evaluate thermal properties of QCD  in the weak-coupling regime. 
In this section, we consider one-loop thermal 
contributions to the static quark propagator, quark-gluon vertices and gluon propagator. 
When the loop momenta and energies are taken at the scale $T$ and the external momenta 
are much lower, so that we may expand with respect to them, these correspond to the hard 
thermal loop (HTL) contributions \cite{Braaten:1989mz}.

\begin{figure}[ht]
\makebox[-5truecm]{\phantom b}
\put(0,0){\epsfxsize=4truecm \epsfbox{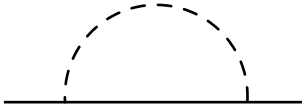}}
\caption {Self-energy diagram. The continuous line stands for a static quark, the dashed 
one for a longitudinal gluon.}
\label{figself}
\end{figure}

\begin{figure}[ht]
\makebox[-5truecm]{\phantom b}
\put(0,0){\epsfxsize=4truecm \epsfbox{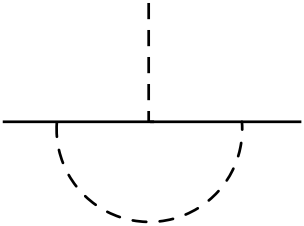}}
\caption {Vertex correction to the static quark line; the incoming gluon is longitudinal.}
\label{figvertexL}
\end{figure}

\begin{figure}[ht]
\makebox[-5truecm]{\phantom b}
\put(0,0){\epsfxsize=4truecm \epsfbox{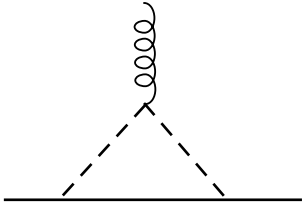}}
\caption {Vertex correction to the static quark line; the incoming gluon is transverse.}
\label{figvertexT}
\end{figure}

The one-loop contributions to the static-quark self energy, to the static-quark lon\-gitudinal-gluon vertex 
and to the  static-quark transverse-gluon vertex are displayed in Figs.~\ref{figself}, 
\ref{figvertexL} and \ref{figvertexT} respectively. It is convenient to fix 
the Coulomb gauge. In that gauge, longitudinal gluons do not depend on the temperature (see Eq. (\ref{D000}))
and the above diagrams do not give thermal contributions. Moreover, if evaluated 
in dimensional regularization they vanish after expanding in the external momenta.
Throughout this work, we will adopt the Coulomb gauge unless stated otherwise.

\begin{figure}[ht]
\makebox[-10truecm]{\phantom b}
\put(0,0){\epsfxsize=10truecm \epsfbox{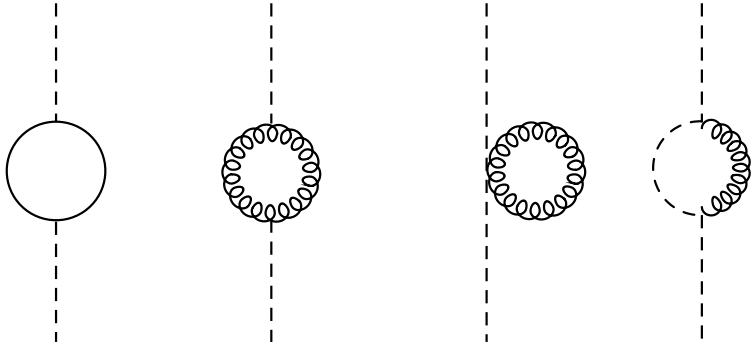}}
\caption{Diagrams contributing to the longitudinal component of the 
gluon polarization tensor at one-loop order. The continuous loop stands for 
light (massless) quark loops, dashed lines for longitudinal gluons and 
curly lines for transverse gluons.
Ghosts do not contribute to the thermal part of the gluon polarization tensor 
\cite{Landshoff:1992ne}.}
\label{figvacpol}
\end{figure}

Momenta and energies of order $T$ contribute to the gluon self-energy diagrams.
Since only longitudinal gluons couple to static quarks, 
we will focus on the longitudinal part of the polarization tensor. This will be the only component 
of the gluon polarization tensor relevant to the paper. Diagrams contributing to the thermal 
part of the longitudinal gluon polarization tensor in real-time formalism at one-loop order 
are shown in Fig.~\ref{figvacpol}. In Sec.~\ref{secvacuumpol}, we will give a general expression, 
and in Secs.~\ref{secvacuumpolk0}, \ref{secvacuumpolk}, \ref{secvacuumpolkoo} 
we will expand it in some relevant limits.

\subsection{The longitudinal gluon polarization tensor}
\label{secvacuumpol}
Summing up all thermal contributions from the diagrams of  Fig.~\ref{figvacpol}, we obtain
(for details see \cite{Ghiglieri08}):
\bea 
\left[\Pi_{00}^{\rm R}(k)\right]_{\rm thermal} &=& 
\left[\Pi_{00}(k^0+i\epsilon,\vec k)\right]_{\rm thermal}\,, 
\label{Pi00fullR} \\
\left[\Pi_{00}^{\rm A}(k)\right]_{\rm thermal} &=& 
\left[\Pi_{00}(k^0-i\epsilon,\vec k)\right]_{\rm thermal}\,, 
\label{Pi00fullA} \\
\nn\\
\left[\Pi_{00}(k)\right]_{\rm thermal} 
&=& \left[\Pi_{00,\,{\rm F}}(k)\right]_{\rm thermal} 
+ \left[\Pi_{00,\,{\rm G}}(k)\right]_{\rm thermal}\,,  
\label{Pi00full} \\
\left[\Pi_{00,\,{\rm F}}(k)\right]_{\rm thermal} 
&=& 
\frac{g^2 \, T_F \, n_f}{2\pi^2}\int_{-\infty}^{+\infty}dq^0\,|q^0|\, n_{\rm F}(\vert q^0\vert)
\nn \\
&& \hspace{-2cm}
\times\left[2-\left(\frac{4(q^0)^2+k^2-4q^0k^0}{4|q^0||{\vec k}|}\right)
\ln\frac{k^2-2q^0k^0+2|q^0||{\vec k}|}{k^2-2q^0k^0-2|q^0||{\vec k}|}\right.
\nn \\
&& \hspace{-2cm} 
\qquad 
\left.
+\left(\frac{4(q^0)^2+k^2+4q^0k^0}{4|q^0||{\vec k}|}\right)
\ln\frac{k^2+2q^0k^0-2|q^0||{\vec k}|}{k^2+2q^0k^0+2|q^0||{\vec k}|}\right]\,,
\label{Pi00fermion}
\\
\left[\Pi_{00,\,{\rm G}}(k)\right]_{\rm thermal} 
&=&
\frac{ g^2 \, N_c}{2\pi^2}\int_{-\infty}^{+\infty}dq^0\,|q^0| n_{\rm B}(|q^0|)
\nn\\
&& \hspace{-2cm} 
\times 
\left\{1+\frac{(2q^0-k^0)^2}{8(q^0)^2}-\frac{1}{2}-\frac{|{\vec k}|^2}{2(q^0)^2}\right.
\nn\\
&& \hspace{-2cm} 
\qquad
+2\left[\frac{|{\vec k}|}{2|q^0|}
-\frac{(|{\vec k}|^2+(q^0)^2)^2}{8|q^0|^3|{\vec k}|}
-\frac{(2q^0-k^0)^2}{4(q^0-k^0)^2}
\left(-\frac{(|{\vec k}|^2+(q^0)^2)^2}{8|q^0|^3|{\vec k}|}
+\frac{|{\vec k}|}{2|q^0|}\right)\right]
\nn\\ 
&& \hspace{-2cm} 
\qquad\qquad\qquad\qquad\qquad\qquad
\times \ln\left\vert\frac{|{\vec k}|-|q^0|}{|{\vec k}|+|q^0|}\right\vert
\nn\\
&& \hspace{-2cm} 
\qquad
-\frac{(2q^0-k^0)^2}{4} \left[\frac{1}{2|q^0||{\vec k}|}+\frac{1}{(q^0-k^0)^2}
\left(\frac{(k^2-2q^0k^0)^2}{8|q^0|^3|{\vec k}|}
+\frac{k^2-2q^0k^0}{2|q^0||{\vec k}|}+\frac{|q^0|}{2|{\vec k}|}\right)\right]
\nn\\
&& \hspace{-2cm} 
\qquad\qquad\qquad\qquad\qquad\qquad
\left.\times\ln\frac{k^2-2q^0k^0+2|q^0||{\vec k}|}{k^2-2q^0k^0-2|q^0||{\vec k}|}\right\}\,,
\label{Pi00gluon}
\eea
where ``R'' stands for retarded, ``A'' for advanced, ``F'' labels the 
contribution coming from the loops of $n_f$ massless quarks (first diagram of Fig.~\ref{figvacpol}) 
and ``G'' labels the contribution from the second, third and fourth diagram of Fig.~\ref{figvacpol}.
$N_c=3$ is the number of colors and $T_F = 1/2$. 
In the context of the imaginary-time formalism, Eqs.~(\ref{Pi00fermion}) 
and (\ref{Pi00gluon}) can be found also in textbooks like \cite{Kapusta:2006pm}. 
The original derivation of (\ref{Pi00gluon}) is in \cite{Heinz:1986kz}.

The retarded and advanced gluon self energies contribute to the retarded 
and advanced gluon propagators. From the retarded and advanced gluon propagators 
we may derive the full propagator, the spectral density and finally all components 
of the $2 \times 2$ matrix of the real-time gluon propagator along the lines 
of Sec.~\ref{secgluonpropagator}. In the following, we study 
Eqs.~(\ref{Pi00fullR})-(\ref{Pi00gluon}) in different kinematical limits.

\subsection{The longitudinal gluon polarization tensor for $k^0 \ll T\sim|\vec k|$}
\label{secvacuumpolk0}
The typical loop momentum $q^0$ is of order $T$. 
If we expand $\left[\Pi_{00}^{\rm R}(k)\right]_{\rm thermal}$ 
and $\left[\Pi_{00}^{\rm A}(k)\right]_{\rm thermal}$ in $k^0 \ll$ $T \sim |\vec k|$ 
and keep terms up to order $k^0$, the result is 
\bea
{\rm Re} \, \left[\Pi_{00}^{\rm R}(k)\right]_{\rm thermal} = 
{\rm Re} \, \left[\Pi_{00}^{\rm A}(k)\right]_{\rm thermal} &=& 
\nn\\
&& \hspace{-7cm}
\frac{g^2 \, T_F \, n_f}{\pi^2} 
\int_0^{+\infty}dq^0\,q^0\,n_{\rm F}(q^0)\left[
2 +  \left( \frac{|\vec k|}{2q^0} - 2 \frac{q^0}{|\vec k|} \right) 
\ln \left| \frac{|\vec k|-2q^0}{|\vec k|+2q^0} \right|
\right]
\nn\\
&& \hspace{-7cm}
+ \frac{g^2 \, N_c}{\pi^2}
\int_0^{+\infty}dq^0\,q^0\,n_{\rm B}(q^0)\left[
1 - \frac{\vec k^2}{2(q^0)^2}
+ \left( - \frac{q^0}{|\vec k|} + \frac{|\vec k|}{2q^0} - \frac{|\vec k|^3}{8(q^0)^3} \right)
\ln \left| \frac{|\vec k|-2q^0}{|\vec k|+2q^0} \right|
\right] \,,
\nn\\
\label{RePi00k0}\\
{\rm Im} \, \left[\Pi_{00}^{\rm R}(k)\right]_{\rm thermal} = 
- {\rm Im} \, \left[\Pi_{00}^{\rm A}(k)\right]_{\rm thermal} &=& 
\nn\\
&& \hspace{-4cm}
\frac{2\, g^2 \, T_F \, n_f}{\pi}\,\frac{k^0}{|\vec k|} 
\int_{|\vec k|/2}^{+\infty}dq^0\,q^0\,n_{\rm F}(q^0)
\nn\\
&& \hspace{-4cm}
+ \frac{g^2 \, N_c}{\pi} \, \frac{k^0}{|\vec k|}
\left[\frac{\vec k^2}{8}\,n_{\rm B}(|\vec k|/2) + 
\int_{|\vec k|/2}^{+\infty}dq^0\,q^0\,n_{\rm B}(q^0)\left( 1 - \frac{\vec k^4}{8(q^0)^4} \right) 
\right]
\,.
\label{ImPi00k0}
\eea
Equation (\ref{ImPi00k0}) and the gluonic part of (\ref{RePi00k0})
are in agreement with \cite{Heinz:1986kz}.

\subsection{The longitudinal gluon polarization tensor for $k^0 \sim |\vec k| \ll T$}
\label{secvacuumpolk}
If we assume that all components of the external four-momentum are much smaller than 
the loop momentum $q^0 \sim T$, then we may expand 
$\left[\Pi_{00}^{\rm R}(k)\right]_{\rm thermal}$ and $\left[\Pi_{00}^{\rm A}(k)\right]_{\rm thermal}$
in $k^0 \sim |\vec k| \ll T$.  At leading order, we obtain  
the well-known HTL expression for the longitudinal gluon polarization tensor, 
which may be found, for instance, in \cite{LeBellac:1996}:
\bea
{\rm Re} \, \left[\Pi_{00}^{\rm R}(k)\right]_{\rm thermal} = 
{\rm Re} \, \left[\Pi_{00}^{\rm A}(k)\right]_{\rm thermal} &=& 
m_D^2 \left( 1- \frac{k^0}{2|\vec k|} \ln \left| \frac{k^0+|\vec k|}{k^0-|\vec k|}\right| \right)\,,
\label{RePi00HTL}\\
{\rm Im} \, \left[\Pi_{00}^{\rm R}(k)\right]_{\rm thermal} = 
- {\rm Im} \, \left[\Pi_{00}^{\rm A}(k)\right]_{\rm thermal} &=& 
m_D^2 \, \frac{k^0}{|\vec k|} \, \frac{\pi}{2} \, \theta(-k^2)\,,
\label{ImPi00HTL}
\eea
where $m_D$ is the Debye mass:  
\be
m_D^2 = \frac{g^2T^2}{3}\left(N_c + T_F\,n_f \right)\,.
\label{mD}
\ee
We have used that $\displaystyle \int_0^\infty dq^0\,q^0\,n_{\rm F}(q^0) = \pi^2 T^2/12$  and 
$\displaystyle \int_0^\infty dq^0\,q^0\,n_{\rm B}(q^0) = \pi^2 T^2/6$. 
Note that the expansions for $|\vec k| \to 0$ of (\ref{RePi00k0}) and (\ref{ImPi00k0}) and 
those for  $k^0 \to 0$ of  (\ref{RePi00HTL}) and (\ref{ImPi00HTL}) agree with each other 
at leading order.

\subsubsection{The resummed longitudinal gluon propagator}
\label{secresummedgluonprop}
The longitudinal polarization tensor induces corrections to the longitudinal gluon propagator:
\be
D_{00}^{\rm R,A}(k) = \frac{i}{\vec k ^2} - \frac{i}{\vec k ^4}  \Pi_{00}^{\rm R,A}(k) + \dots \;.
\label{D00HTLexp}
\ee
Since $\Pi_{00}^{\rm R,A}$ contains a real and an imaginary part, also $D_{00}^{\rm R,A}$ 
acquires a real and an imaginary part.

If the typical momentum transfer is of the order of the Debye mass, $|\vec k| \sim m_D$,
then the series in (\ref{D00HTLexp}) needs to be resummed:
\be
D_{00}^{\rm R,A}(k) = \frac{i}{\vec k ^2 + \Pi_{00}^{\rm R,A}(k)} \,.
\label{D00HTLresum}
\ee
The resummed longitudinal propagator depends on $k^0$ and has a 
real and an imaginary part. The Debye mass $m_D$ plays the role of a screening mass  for 
longitudinal gluons whose momenta are such that $k^0 \ll T$ and $|\vec k| \sim m_D$.
A study of the resummed gluon propagator in the real-time formalism 
may be found in \cite{Carrington:1997sq}.

The role of the screening mass can be made more evident if we assume further 
that $k^0 \ll |\vec k| \sim m_D \ll T$ and expand Eq. (\ref{D00HTLresum}) in $k^0$
up to order $k^0$; then we obtain 
\be
D_{00}^{\rm R,A}(k) = \frac{i}{\vec k ^2 + m_D^2}  
\pm  \frac{\pi}{2}\,\frac{k^0}{|\vec k|}\,\frac{m_D^2}{\left(\vec k ^2 + m_D^2\right)^2} \,,
\label{D00RAHTLresumk0}
\ee
where the ``$+$'' and ``$-$'' signs refer to the retarded and advanced propagators respectively.
The corresponding spectral density is
\be
\rho_{\rm B\, 00}(k) = D_{00}^{\rm R}(k) - D_{00}^{\rm A}(k) = 
\pi\,\frac{k^0}{|\vec k|}\,\frac{m_D^2}{\left(\vec k ^2 + m_D^2\right)^2} 
\,.
\ee
Following Sec.~\ref{secgluonpropagator} and expanding in $k^0$ also the 
Bose factor, $n_{\rm B}(k^0) \approx T/k^0$, at leading order in $k^0$, we obtain 
that the resummed HTL longitudinal propagator in the real-time formalism is 
\bea
{\bf D}_{00}(0,\vec k) &=& 
 \frac{i}{\vec k ^2 + m_D^2}  
\left(
\begin{matrix}
&&\hspace{-2mm}1
&&0 \\ 
&&\hspace{-2mm}0
&&-1
\end{matrix}
\right) 
+
\pi \,\frac{T}{|\vec k|}\,\frac{m_D^2}{\left(\vec k ^2 + m_D^2\right)^2}
\left(
\begin{matrix}
&&\hspace{-2mm}1
&&1 \\ 
&&\hspace{-2mm}1
&&1
\end{matrix}
\right) 
\,.
\label{D00HTLresumk0}
\eea

\subsection{The longitudinal gluon polarization tensor for $|\vec k| \gg T \gg k^0$}
\label{secvacuumpolkoo}
If we assume that  $|\vec k| \gg T \gg k^0$, then the expression for the longitudinal 
gluon polarization tensor may extracted from Eqs.~(\ref{RePi00k0}) and (\ref{ImPi00k0})
by expanding for large $|\vec k|/T$. At leading order, we obtain 
\be
\left[\Pi_{00}^{\rm R}(k)\right]_{\rm thermal} = \left[\Pi_{00}^{\rm A}(k)\right]_{\rm thermal}
= - \frac{N_c\,g^2\,T^2}{18}\,.
\label{Pi00koo}
\ee 
The result is real and does not depend on $k$. Moreover, only the gluonic part of 
the polarization tensor contributes in this limit and at this order.
Higher-order real corrections are suppressed by $T^2/\vec k^2$, while higher-order imaginary 
corrections are exponentially suppressed.

\section{Bound states for $1/r \gg T$}
\label{secWpNRQCD}
Starting from this section, we shall address bound states made of 
a static quark and antiquark in QCD at finite $T$. Bound states 
introduce extra scales in the dynamics, besides $T$ and $m_D$, 
that we have to account for. The most relevant one is the distance $r$ between 
the quark and the antiquark. Throughout  the paper, we will assume that 
$1/r \gg \lQ$. We will further assume that also the binding energy of the 
quark-antiquark static pair is larger than $\lQ$. 

First, we deal with the situation where the inverse distance 
of the two static sources is much larger than the temperature: $1/r \gg T$. 
Under this condition, we may integrate out $1/r$ from static QCD at $T=0$ order 
by order in $\als$. The EFT that we obtain is potential non-relativistic QCD (pNRQCD) 
in the static limit \cite{Pineda:1997bj,Brambilla:1999xf}, 
whose Lagrangian can be written as 
\bea
{\cal L} = 
&& 
- \frac{1}{4} F^a_{\mu \nu} F^{a\,\mu \nu} 
+ \sum_{i=1}^{n_f}\bar{q}_i\,iD\!\!\!\!/\,q_i 
+ \int d^3r \; {\rm Tr} \,  
\Biggl\{ {\rm S}^\dagger \left[ i\partial_0 + C_F\frac{\alVs}{r} \right] {\rm S} 
+ {\rm O}^\dagger \left[ iD_0 - \frac{1}{2N_c}\frac{\alVo}{r} \right] {\rm O} \Biggr\}
\nn\\
&& 
+ V_A\, {\rm Tr} \left\{  {\rm O}^\dagger {\vec r} \cdot g{\vec E} \,{\rm S}
+ {\rm S}^\dagger {\vec r} \cdot g{\vec E} \,{\rm O} \right\} 
+ \frac{V_B}{2} {\rm Tr} \left\{  {\rm O}^\dagger {\vec r} \cdot g{\vec E} \, {\rm O} 
+ {\rm O}^\dagger {\rm O} {\vec r} \cdot g{\vec E}  \right\}  + \dots\;.
\label{pNRQCD}
\eea
The fields ${\rm S} = S \, {1\!\!{\rm l}_c / \sqrt{N_c}}$ and ${\rm O} =  O^a\, {T^a / \sqrt{T_F}}$ 
are static quark-antiquark singlet and octet fields respectively, 
$\vec E$ is the chromoelectric field: $E^i = F^{i0}$, and  $C_F = (N_c^2-1)/(2 N_c) = 4/3$. 
The trace is over the color indeces.
The matching coefficients $\alVs$, $\alVo$, $V_A$, $V_B$ are at leading order:
$\alVs = \als$, $\alVo = \als$, $V_A = 1$ and $V_B = 1$.
Gluon fields are multipole expanded and depend only on the center of mass coordinate;
they scale with the low-energy scales ($T$, $m_D$, $\als/r$, $\lQ$, ...) 
that are still dynamical in the EFT. 
The dots in the last line stand for higher-order terms in the multipole expansion.

\subsection{Singlet and octet propagators}
\label{secsingletoctetrT}
The free real-time singlet and octet static propagators at finite $T$ are similar 
to the free static quark propagator (\ref{Squarkstatic}), although singlet and octet are bosons:
\bea
{\bf S}^{\rm singlet}(p) &=& 
\left(
\begin{matrix}
&&\hspace{-2mm}\displaystyle\frac{i}{p^0 + C_F \, \alVs/r +i\epsilon}  
&&0 \\ 
&&\hspace{-2mm}2\pi\delta(p^0 + C_F \, \alVs/r)  
&&\displaystyle\frac{-i}{p^0 + C_F\, \alVs/r -i\epsilon}
\end{matrix}
\right)
\nn\\
&=& 
{\bf U}^{(0)} 
\left(
\begin{matrix}
&&\hspace{-2mm}\displaystyle\frac{i}{p^0 + C_F \, \alVs/r +i\epsilon} 
&&0 \\ 
&&0  
&&\displaystyle\frac{-i}{p^0 + C_F\, \alVs/r -i\epsilon}
\end{matrix}
\right)
{\bf U}^{(0)}\,,
\label{Ssingletstatic}
\eea
\bea
{\bf S}^{\rm octet}(p)_{ab} &=& 
\delta_{ab}\,\left(
\begin{matrix}
&&\hspace{-2mm}\displaystyle\frac{i}{p^0 - \alVo/(2N_c\, r) +i\epsilon}
&&0 \\ 
&&\hspace{-2mm}2\pi\delta(p^0 - \alVo/(2N_c\, r))
&&\displaystyle\frac{-i}{p^0 - \alVo/(2N_c\, r) - i\epsilon}
\end{matrix}
\right)
\nn\\
&=& 
\delta_{ab}\,{\bf U}^{(0)} 
\left(
\begin{matrix}
&&\hspace{-2mm}\displaystyle\frac{i}{p^0 - \alVo/(2N_c\, r) +i\epsilon}
&&0 \\ 
&&0  
&&\displaystyle\frac{-i}{p^0 - \alVo/(2N_c\, r) -i\epsilon}
\end{matrix}
\right)
{\bf U}^{(0)}\,.
\label{Soctetstatic}
\eea

\subsection{Non-thermal part of the singlet static potential}
The contribution to the singlet static potential coming from the scale $1/r$ can be read 
from the Lagrangian (\ref{pNRQCD}); it is just the Coulomb potential, which 
in real-time formalism reads 
\be 
{\bf V_s}(r) =  -C_F \frac{\alVs(1/r)}{r}
\left(
\begin{matrix}
&&\hspace{-2mm}1
&&0 \\ 
&&\hspace{-2mm}0
&&-1
\end{matrix}
\right),
\label{Vssoft}
\ee
where $\alVs(1/r)$ is a series in $\als$: at leading order, $\alVs(1/r) = \als(1/r)$.
Starting from order $\als^4$,  $\alVs$ is infrared divergent; these divergences, which 
appear at zero temperature, have been considered elsewhere \cite{Brambilla:1999qa}
and will not matter here.

The matrix in (\ref{Vssoft}) is such that
\be
\left(
\begin{matrix}
&&\hspace{-2mm}1
&&0 \\ 
&&\hspace{-2mm}0
&&-1
\end{matrix}
\right)
= [{\bf U}^{(0)}]^{-1}
\left(
\begin{matrix}
&&\hspace{-2mm}1
&&0 \\ 
&&\hspace{-2mm}0
&&-1
\end{matrix}
\right)
[{\bf U}^{(0)}]^{-1}\,.
\label{diag1}
\ee
Together with Eq.~(\ref{diagS0}), this guarantees that 
\be
{\bf S}^{\rm singlet}(p) = 
\left(
\begin{matrix}
&&\hspace{-2mm}\displaystyle\frac{i}{p^0 +i\epsilon}  
&&0 \\ 
&&\hspace{-2mm}2\pi\delta(p^0) 
&&\displaystyle\frac{-i}{p^0 -i\epsilon}
\end{matrix}
\right)
\sum_{n=0}^\infty 
\left[
\left(-i {\bf V_s}(r)\right)
\left(
\begin{matrix}
&&\hspace{-2mm}\displaystyle\frac{i}{p^0 +i\epsilon}  
&&0 \\ 
&&\hspace{-2mm}2\pi\delta(p^0) 
&&\displaystyle\frac{-i}{p^0 -i\epsilon}
\end{matrix}
\right)
\right]^n,
\label{sumSinglet}
\ee
i.e. that, like at $T=0$, the sum of all insertions of a potential exchange between 
a free quark and antiquark amounts to the propagator (\ref{Ssingletstatic}).

The singlet static potential does not get only contributions from the scale $1/r$, 
but, if the next relevant scales are the thermal ones, it will also get thermal 
contributions. These will be the subject of the following sections.

\subsection{Chromoelectric correlator}
\label{secEE}
In the paper, it will become necessary to calculate the chromoelectric correlator 
$\langle \vec E^a(t)\phi(t,0)^{\rm adj}_{ab} \vec E^b(0)\rangle_T$
where $\phi(t,0)^{\rm adj}_{ab}$ is a Wilson line in the adjoint representation connecting 
the points $(t,\vec 0)$ and $(0,\vec 0)$ by a straight line. Such a correlator 
enters each time we consider diagrams with two chromoelectric dipole insertions. 
In the real-time formalism, the chromoelectric correlator is a $2\times 2$ 
matrix in the field indices ``1'' and ``2''. We shall write it as 
\be 
\langle \vec E^a(t)\phi(t,0)^{\rm adj}_{ab} \vec E^b(0)\rangle_T  
= (N_c^2-1) \, \int \frac{dk^0}{2\pi} e^{-ik^0t} \int \frac{d^3k}{(2\pi)^3} \left(
(k^0)^2 \, {\bf D}_{ii}(k) + \vec k^2 \, {\bf D}_{00}(k) 
\right)\,,
\ee
where, at zeroth order in $\als$, ${\bf D}_{\mu\nu}(k)$ is the free gluon propagator. 
In Coulomb gauge, the free gluon propagator has been given in Eqs. (\ref{D000}) and (\ref{D0ij});   
since the chromoelectric correlator is a gauge invariant quantity the choice of the gauge does not matter. 
At zeroth order in $\als$, the thermal part of 
$\langle \vec E^a(t)\phi(t,0)^{\rm adj}_{ab} \vec E^b(0)\rangle_T$ is 
\be
\left. \langle \vec E^a(t)\phi(t,0)^{\rm adj}_{ab} \vec E^b(0)\rangle_T  \right|_{\hbox{thermal part}}
= (N_c^2-1) \, \int \frac{d^3k}{(2\pi)^3} \; 2 |\vec k| \cos (|\vec k|t) \, n_{\rm B}(|\vec k|) \,  
\left(
\begin{matrix}
&&\hspace{-2mm}1
&&1 \\ 
&&\hspace{-2mm}1  
&&~1
\end{matrix}
\right)\,.
\label{EEtT}
\ee
Note that for $t=0$ Eq. (\ref{EEtT}) gives the thermal part of the gluon condensate 
in the weak-coupling regime:
\be
\left. \langle \vec E^a(0)\cdot \vec E^a(0)\rangle_T \right|_{\hbox{thermal part}}
= (N_c^2-1) \, T^4 \, \frac{\pi^2}{15}
\left(
\begin{matrix}
&&\hspace{-2mm}1
&&1 \\ 
&&\hspace{-2mm}1  
&&~1
\end{matrix}
\right)
\,,
\label{StefanBoltzmann}
\ee
where we have used $\displaystyle \int_0^\infty dk \,k^3\,n_{\rm B}(k) = \pi^4 T^4/15$.
Equation (\ref{StefanBoltzmann}) agrees with the Stefan--Boltzmann law 
(see, for instance, \cite{Schaefer:1998wd}).

\subsection{Thermal corrections to the singlet static potential}
We calculate now the leading thermal contributions to the static  
potential assuming for definiteness that $T$ and $m_D$ are the next relevant scales after $1/r$, 
i.e. that the binding energy is much smaller than $m_D$.
We will further assume that all other thermodynamical scales are much smaller than the 
binding energy, so that we can ignore them.
We recall that from an EFT point of view, only energy scales larger than the binding energy contribute 
to the potential, which is the matching coefficient entering the Schr\"odinger equation of the bound state, 
while all energy scales contribute to physical observables like, for instance, 
the static energy \cite{Brambilla:2004jw}.
We will comment on the impact of degrees of freedom with energies and momenta 
of the order of the binding energy in Secs.~\ref{secstaticenergy} and \ref{commentI}.

The calculation proceeds in two steps. The first step will be performed in Sec.~\ref{secshortT}.
It consists in integrating out from the pNRQCD Lagrangian (\ref{pNRQCD}) modes of energy and momentum of order $T$.
This modifies pNRQCD into a new EFT where only modes with energy and momentum lower than the temperature  
are dynamical. For our purposes, it only matters that 
the pNRQCD Lagrangian gets additional contributions in the singlet and in the Yang--Mills sectors.
In the first sector, it is an additional contribution to the singlet static potential.
In the second one, the additional contribution corresponds to the HTL Lagrangian \cite{Braaten:1991gm}.

The second step, which will be performed in Sec.~\ref{secshortmD}, consists in integrating out from 
the previous EFT modes of energy and momentum of order $m_D$. The resulting EFT will only have degrees of freedom 
that are dynamical at energy and momentum scales lower than the Debye mass. In the singlet sector, 
the EFT gets modified by a further additional contribution to the static potential.

In summary, if both $T$ and $m_D$ are much larger than the binding energy, 
the real-time Coulomb potential (\ref{Vssoft}), gets two type of corrections 
$\delta {\bf V_s}(r)$. The first one comes from the scale $T$ and the other one from the scale $m_D$.

\subsubsection{Contributions from the scale $T$}
\label{secshortT}
\begin{figure}[ht]
\makebox[-7truecm]{\phantom b}
\put(0,0){\epsfxsize=7truecm \epsfbox{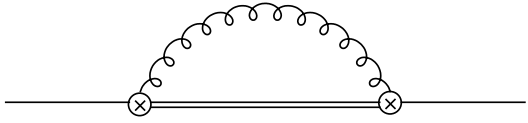}}
\caption {The single continuous line stands for a singlet propagator, 
the double line for an octet propagator, the circle with a cross for a 
chromoelectric dipole vertex and the curly line connecting the two circles with a 
cross for a chromoelectric correlator.}
\label{figEE}
\end{figure}

The leading thermal correction $\delta {\bf V}_s(r)$ is induced by the dipole terms 
${\rm O}^\dagger {\vec r} \cdot g{\vec E} \,{\rm S}$ $+$  
${\rm S}^\dagger {\vec r} \cdot g{\vec E} \,{\rm O}$ in the Lagrangian (\ref{pNRQCD}).
It reads (see \cite{Brambilla:1999xf} for the $T=0$ case):
\bea
\left[ \delta {\bf V_s}(r)\right]_{11} &=& 
- i g^2 \, \frac{T_F}{N_c} \, \frac{r^2}{d-1}
\int_0^\infty \! dt \, e^{-it\Delta V} 
\left[\langle \vec E^a(t)\phi(t,0)^{\rm adj}_{ab} \vec E^b(0)\rangle_T\right]_{11}
 \nn\\
&=& 
- i g^2 \, C_F \, \frac{r^2}{d-1}
\mu^{4-d} \!\int \!\! \frac{d^dk}{(2\pi)^d}
\frac{i}{-k^0 -\Delta V +i\epsilon}
\left[(k^0)^2 \, {\bf D}_{ii}(k) + \vec k^2 \, {\bf D}_{00}(k) 
\right]_{11}\!,
\label{EETfull11}
\\
\left[ \delta {\bf V_s}(r)\right]_{22} &=& - \left[ \delta {\bf V_s}\right]^*_{11} \,,
\label{EETfull22}
\\
\left[ \delta {\bf V_s}(r)\right]_{12} &=& 0\,,
\label{EETfull12}
\\
\left[ \delta {\bf V_s}(r)\right]_{21} &=& 
i g^2 \, C_F \, \frac{r^2}{d-1}
\mu^{4-d} \! \int \!\! \frac{d^dk}{(2\pi)^d}
\, 2\pi \delta(-k^0-\Delta V) \, 
\left[(k^0)^2 \, {\bf D}_{ii}(k) + \vec k^2 \, {\bf D}_{00}(k) 
\right]_{21},
\label{EETfull21}
\eea
where $\displaystyle \Delta V = \frac{1}{r}\left(\frac{\alVo}{2N_c}+ C_F \alVs\right) \approx \frac{N_c\als}{2r}$.
The corresponding Feynman diagram is shown in Fig.~\ref{figEE}. 
Integrals over momenta have been regularized in dimensional regularization
($d$ is the number of dimensions, $\mu$ is the compensating scale).
In Eq.~(\ref{EETfull11}), $i/(-k^0 -\Delta V +i\epsilon)$ is the ``11'' component of the static  
octet propagator; Eq.~(\ref{EETfull12}) vanishes because the ``12'' component of the static octet 
propagator vanishes and in  Eq.~(\ref{EETfull21}), $ 2\pi \delta(-k^0-\Delta V)$ is the 
``21'' component of the static  octet propagator. Note that vertices of type ``1'' and ``2'' have 
opposite signs. Equation (\ref{EETfull22}), which may also be read  
$\left[-i\delta {\bf V_s}(r)\right]_{22} = \left[-i\delta {\bf V_s}(r)\right]^*_{11}$,  
reflects the relation existing between the ``11'' and ``22'' components of the propagators in the 
real-time formalism.

We are interested in calculating the contribution to the integrals 
in Eqs.~(\ref{EETfull11})-(\ref{EETfull21}) from momenta $k \sim T$.
Since $T \gg \Delta V$, we may expand in $\Delta V/T$. 
Moreover, at leading order, the propagators in  Eqs.~(\ref{EETfull11}) and (\ref{EETfull21}) 
are the free ones, ${\bf D}^{(0)}_{00}$ and ${\bf D}^{(0)}_{ii}$, 
given in Eqs.~(\ref{D000}) and (\ref{D0ij}). However the leading-order thermal contribution, 
which would be of order $g^2\, r^2\, T^3$, vanishes:  
\bea 
\left[ \delta {\bf V_s}(r)\right]_{11} &=& 
- i g^2 \, C_F \, \frac{r^2}{d-1}
\mu^{4-d} \int \frac{d^dk}{(2\pi)^d}
\frac{i}{-k^0 +i\epsilon}
\, (k^0)^2 \, \, 4\pi \delta(k^2) \, n_{\rm B}(|k^0|)  = 0 \,,  
\label{leadingT11}\\
\left[ \delta {\bf V_s}(r)\right]_{21} &=& 
i g^2 \, C_F \, \frac{r^2}{d-1}
\mu^{4-d} \int \frac{d^dk}{(2\pi)^d}
\, 2\pi \delta(-k^0) 
\, (k^0)^2 \, \, 4\pi \delta(k^2) \, n_{\rm B}(|k^0|)  = 0 \,.
\label{leadingT21}
\eea

Several next-to-leading order corrections are possible, because 
several scales are still dynamical in the EFT: we may have 
corrections of relative order $\Delta V/T$, $m_D/T$, $(r T)$, $\als$ and so on.

{\bf (1)} First, we consider corrections of order $\Delta V/k^0$ or higher 
to the quark-antiquark propagator, which contribute to order  
$g^2\, r^2\, T^3 \times \Delta V/T$ or higher to  $\delta {\bf V_s}(r)$:
\bea
\left[ \delta {\bf V_s}(r)\right]_{11} &=& 
- i g^2 \, C_F \, \frac{r^2}{d-1}
\mu^{4-d} \int \frac{d^dk}{(2\pi)^d}
\frac{i}{-k^0 -\Delta V +i\epsilon}
\left[(k^0)^2 \, {\bf D}^{(0)}_{ii}(k) + \vec k^2 \, {\bf D}^{(0)}_{00}(k) 
\right]_{11}
\nn\\
&=&
\frac{4}{3}\, C_F \, \frac{\als}{\pi} \, r^2 \, T^2\, \Delta V\, 
f\!\left({\Delta V}/{T} \right)
- i \frac{2}{3}\, C_F \, \als\, r^2\, (\Delta V)^3 \, n_{\rm B}(\Delta V)\,,
\label{dVT11}
\\
\left[ \delta {\bf V_s}(r)\right]_{21} &=& 
i g^2 \, C_F \, \frac{r^2}{d-1}
\mu^{4-d} \int \frac{d^dk}{(2\pi)^d}
\, 2\pi \delta(-k^0-\Delta V) \, 
\left[(k^0)^2 \, {\bf D}^{(0)}_{ii}(k) + \vec k^2 \, {\bf D}^{(0)}_{00}(k) 
\right]_{21}
\nn\\
&=&
 i \frac{4}{3}\, C_F \, \als\, r^2\, (\Delta V)^3 \, n_{\rm B}(\Delta V)\,
= -2\,i\, {\rm Im}\, \left[ \delta {\bf V_s}(r)\right]_{11} \,,
\label{dVT21}
\eea
where 
\be
f(z) = \int_0^\infty dx\, \frac{x^3}{e^x-1} \; {\rm P} \frac{1}{x^2-z^2}\,,
\label{ffunction}
\ee
and  P stands for the principal value.
Since $T \gg \Delta V$, we can expand the above expressions in $\Delta V/T$ obtaining
\bea
\delta {\bf V_s}(r)
&=&  \frac{\pi}{9} \, N_c C_F \, \als^2 \, r \, T^2\, 
\left(
\begin{matrix}
&&\hspace{-2mm}1
&&0 \\ 
&&\hspace{-2mm}0
&&-1
\end{matrix}
\right) + \dots 
- \frac{i}{6} \, N_c^2 C_F \, \als^3\, T\,  
\left(
\begin{matrix}
&&\hspace{-2mm}1
&&0 \\ 
&&\hspace{-2mm}-2
&&1
\end{matrix}
\right) + \dots 
\,.
\label{VsT}
\eea
The dots stand for higher-order real and imaginary terms.
We note that the matrices in Eq.~(\ref{VsT}) 
are such to be diagonalized by the matrix $[{\bf U}^{(0)}]^{-1}$: 
the first one is the same as in (\ref{diag1}), the second one satisfies
\be
\left(
\begin{matrix}
&&\hspace{-2mm}1
&&0 \\ 
&&\hspace{-2mm}-2 
&&1
\end{matrix}
\right)
= [{\bf U}^{(0)}]^{-1}
\left(
\begin{matrix}
&&\hspace{-2mm}1
&&0 \\ 
&&\hspace{-2mm}0
&&1
\end{matrix}
\right)
[{\bf U}^{(0)}]^{-1}\,.
\ee

\begin{figure}[ht]
\makebox[-7truecm]{\phantom b}
\put(0,0){\epsfxsize=7truecm \epsfbox{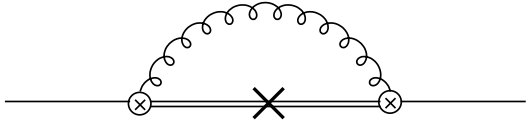}}
\caption {Feynman diagram giving the leading order real contribution to Eq.~(\ref{dVT11}).
The symbols are like in Fig.~\ref{figEE}. The cross stands for a $\Delta V$ insertion in the octet propagator.}
\label{figEEV}
\end{figure}

\begin{figure}[ht]
\makebox[-7truecm]{\phantom b}
\put(0,0){\epsfxsize=7truecm \epsfbox{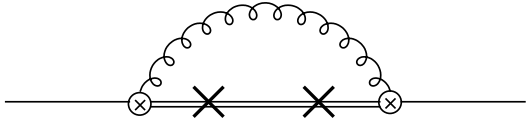}}
\caption {Feynman diagram giving the leading order imaginary contribution to Eq.~(\ref{dVT11}).
The symbols are like in Fig.~\ref{figEEV}.}
\label{figEEV2}
\end{figure}

The leading real contribution in Eq.~(\ref{VsT}) is of order $g^2\, r^2\, T^3 \times \Delta V/T$
and comes from the diagram shown in Fig.~\ref{figEEV}. The leading imaginary contribution in Eq.~(\ref{VsT}) 
is of order $g^2\, r^2\, T^3 \times (\Delta V/T)^2$ and comes from the diagram shown in Fig.~\ref{figEEV2}.
This imaginary part is a very peculiar feature of QCD. It originates from the fact that thermal fluctuations 
of the medium at short distances may destroy a color-singlet bound state in an octet quark-antiquark state 
and gluons. This process is kinematically not allowed at zero temperature where only static octets
may decay into singlets with a leading order width $\Gamma = N_c^2\,\als^4/(12r)$.
In some kinematical situations, the role of singlet-octet transitions in quarkonium-hadron scattering 
in strongly interacting matter has been discussed in Ref.~\cite{Kharzeev:1994pz}.

Here we have assumed $T \gg \Delta V$. In the kinematical situation 
$T \sim \Delta V$, Eq.~(\ref{dVT11}) would provide a contribution to the 
static energy rather than to the static potential of the quark-antiquark pair.  
This contribution will be discussed in Sec.~\ref{secstaticenergy}.

\begin{figure}[ht]
\makebox[-7truecm]{\phantom b}
\put(0,0){\epsfxsize=7truecm \epsfbox{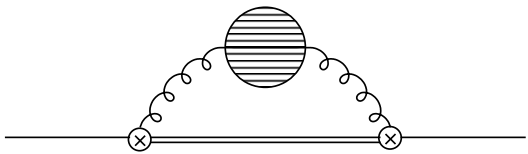}}
\caption {The symbols are like in Fig.~\ref{figEE}. The dashed blob 
stands for a one-loop self-energy insertion in the gluon propagator.}  
\label{figEEmD}
\end{figure}

{\bf (2)} Another source of next-to-leading order corrections comes from next-to-leading order 
corrections to the chromoelectric field correlator. Their contribution to the static potential 
at zero temperature has been considered in \cite{Brambilla:2006wp}.
They also contribute at order  $g^2\, r^2\, T^3 \times \als$ 
$ \sim g^2 \, r^2 \, T^3 \times (m_D/T)^2$ to  the thermal part of the static potential. 
The chromoelectric correlator enters in the potential in the expression 
$\displaystyle \mu^{4-d} \int \frac{d^dk}{(2\pi)^d} \, \frac{i}{-k^0+i\epsilon} \,  
\left[(k^0)^2 \, {\bf D}_{ii}(k) + \vec k^2 \, {\bf D}_{00}(k) \right]_{11}$.
Since $i/(-k^0 +i\epsilon) = -i\, {\rm P}\, (1/k^0) + \pi \delta(-k^0)$ and 
$\left[(k^0)^2 \, {\bf D}_{ii}(k) + \vec k^2 \, {\bf D}_{00}(k) \right]_{11}$ is even 
in $k^0$, only the $\pi \delta(-k^0)$ component of the static quark-antiquark propagator
contributes, therefore only the limit for $k^0\to 0$  of 
$\left[(k^0)^2 \, {\bf D}_{ii}(k) + \vec k^2 \, {\bf D}_{00}(k)\right]$ matters.
In order to evaluate it, it is convenient to perform the calculation first in temporal-axial gauge 
$A^0 =0$. In this gauge, the chromoelectric field is simply $\vec E = - \partial_0 \vec A$. Hence  
all corrections to the chromoelectric correlator are encoded in the spatial part of the gluon propagator alone: 
at one loop the correction is provided entirely by the gluon self energy. 
In temporal-axial gauge, from the transversality relation of the polarization tensor 
it follows that (compare with the expressions of the propagators in \cite{Heinz:1986kz}):
\be 
\lim_{k^0\to 0} \; (k^0)^2 D_{ii}^{\rm R,A}(k)\bigg|_{\rm temporal-axial \; gauge} = 
\lim_{k^0\to 0} \; i \frac{\vec k^2}{\vec k^2 + \Pi_{00}^{\rm R,A}(k)}\bigg|_{\rm temporal-axial \; gauge} \,.
\ee
Since in the $k^0\to 0$ limit $\Pi_{00}^{\rm R,A}(k)$ is equal in Coulomb and temporal-axial gauge, 
we can also write that (compare with Eq.~(\ref{D00HTLresum}))
\be 
\lim_{k^0\to 0} \; (k^0)^2 D_{ii}^{\rm R,A}(k)\bigg|_{\rm temporal-axial \; gauge} = 
\lim_{k^0\to 0} \; \vec k^2 D_{00}^{\rm R,A}(k)\bigg|_{\rm Coulomb \; gauge}\,.
\ee
The left-hand side is the only term of the chromoelectric correlator contributing to the potential 
in temporal-axial gauge: it may be evaluated by calculating the right-hand side in Coulomb gauge. 
At one loop, the right-hand side gets contribution from the gluon self-energy diagram 
shown in Fig.~\ref{figEEmD}; hence, at next-to-leading order we can write 
\bea
\left[ \delta {\bf V_s}(r)\right]_{11} &=&
- i g^2 \, C_F \, \frac{r^2}{d-1}
\mu^{4-d} \int \frac{d^dk}{(2\pi)^d} 
\, \pi \delta(-k^0) \,  
\vec k^2 \, \left[\delta{\bf D}_{00}(k) \right]_{11}\,,
\label{Vs11Teq1}\\
\left[ \delta {\bf V_s}(r)\right]_{21} &=& 
i g^2 \, C_F \, \frac{r^2}{d-1}
\mu^{4-d} \int \frac{d^dk}{(2\pi)^d}
\, 2\pi \delta(-k^0) \,
\vec k^2 \left[\delta{\bf D}_{00}(k) \right]_{21} \,,
\label{Vs21Teq2}
\eea
where 
\bea 
\left[\delta{\bf D}_{00}(k) \right]_{11} &=&  \frac{\delta D_{00}^{\rm R}(k)+ \delta D_{00}^{\rm A}(k)}{2}  
+ \left(\frac{1}{2} + n_{\rm B}(k^0)\right) \left(\delta D_{00}^{\rm R}(k) - \delta D_{00}^{\rm A}(k) \right)\,, 
\label{D0011}
\\
\left[\delta{\bf D}_{00}(k) \right]_{21} &=&  (1+n_{\rm B}(k^0))
\left(\delta D_{00}^{\rm R}(k) - \delta D_{00}^{\rm A}(k) \right)\,, 
\label{D0021}\\
\delta D_{00}^{\rm R,A}(k) &=& - \frac{i}{\vec k ^4}  \Pi_{00}^{\rm R,A}(k) \,,
\label{DRAPi}
\eea
and, the relevant limit for the gluon polarization $\Pi_{00}^{\rm R,A}(k)$ in Coulomb gauge 
is given by Eqs.~(\ref{RePi00k0}) and (\ref{ImPi00k0}).
Finally, the correction to the real-time potential reads 
\bea
\delta {\bf V_s}(r)
&=& 
\left[ - \frac{3}{2} \zeta(3)\,  C_F \, \frac{\als}{\pi} \, r^2 \, T \,m_D^2
+ \frac{2}{3} \zeta(3)\, N_c C_F \, \als^2 \, r^2 \, T^3 \right]
\left(
\begin{matrix}
&&\hspace{-2mm}1
&&0 \\ 
&&\hspace{-2mm}0
&&-1
\end{matrix}
\right)
\nn\\
&&
+ i \left[ \frac{C_F}{6} \als \, r^2 \, T \,m_D^2\, \left( 
\frac{1}{\epsilon} + \gamma_E + \ln\pi 
- \ln\frac{T^2}{\mu^2} + \frac{2}{3} - 4 \ln 2 - 2 \frac{\zeta^\prime(2)}{\zeta(2)} \right)
\right.
\nn\\
&& \quad \left.
+ \frac{4\pi}{9} \ln 2 \; N_c C_F \,  \als^2\, r^2 \, T^3
\right]
\left(
\begin{matrix}
&&\hspace{-2mm}1
&&0 \\ 
&&\hspace{-2mm}-2
&&1
\end{matrix}
\right)
\,,
\label{VsTloop}
\eea
where $\epsilon = (4-d)/2$, $\gamma_E$ is the Euler gamma and $\zeta$ 
the Riemann zeta function ($\zeta(2) = \pi^2/6$).
Note that in Eq.~(\ref{VsTloop}), besides terms that are  proportional to the Debye mass there are finite terms,
both in the real and in the imaginary part, that do not depend on it.

Equation (\ref{VsTloop}) contains an imaginary contribution. 
The origin of this contribution is different from the one in Eq.~(\ref{VsT}).
The one here comes from the imaginary part in the gluon self energy, which is due to 
to the scattering of particles with momenta of order $T$ in the thermal bath 
with space-like gluons, $(k^0)^2 <|\vec{k}|^2$, (Landau damping)  while the one in Eq.~(\ref{VsT}) 
signals the thermal break up of a quark-antiquark color singlet pair into an octet one.

The result in Eq.~(\ref{VsTloop}) is infrared divergent and, in an EFT language, calls 
for an opposite ultraviolet divergence from lower energy contributions: the two divergences 
should cancel in all physical observables. In the following section, 
we will show that the infrared divergence generated by the diagram in Fig.~\ref{figEEmD} 
when integrated over momenta of order $T$ is canceled by an ultraviolet divergence 
in the same diagram when integrated over momenta of order $m_D$. 
For the purpose of the cancellation and the calculation of the static 
potential in the situation $1/r \gg T \gg m_D \gg \Delta V$ 
it is irrelevant how we regularize both divergent contributions, as long as they are regularized 
in the same way: the divergences as well as any scheme 
dependence cancel in the sum. However, it may be that expression (\ref{VsTloop}) is used 
in intermediate calculations that require further regularizations.
An example is the solution of the Schr\"odinger equation for quarkonium in a thermal medium when 
the kinetic energy of the bound state is larger than or of the same order as $m_D$ so that the cancellation 
of the divergences does not occur at the level of the potential. In such a situation, it is important 
to provide divergent contributions in a standard regularization scheme.  Equation (\ref{VsTloop}) has 
been obtained by regularizing in dimensional regularization only the integral in $k$.  
This is not sufficient to qualify Eq.~(\ref{VsTloop}) as a standard dimensional regularization 
scheme result, which would also require the calculation of the gluon polarization tensor 
at order $\epsilon$ in the dimensional regularization expansion. This order, combined with the $1/\epsilon$ 
divergence of the integral in $k$, contributes to the finite part of (\ref{VsTloop}).
To our knowledge the expression of the gluon polarization tensor at order $\epsilon$ is not known 
in the literature and its calculation is beyond the purposes of this work. However, it may become 
necessary for a proper calculation of the quarkonium potential at short distances 
when the Debye mass is of the same order as or smaller than the kinetic energy. 

{\bf (3)} Finally, a source of higher-order contributions comes from higher-order 
terms in the multipole expansion. These are of two types. First, they may involve 
operators of higher order in the multipole expansion in the pNRQCD Lagrangian, like, 
for example, the operator $\displaystyle \frac{1}{24} {\rm Tr} \left\{\! {\rm O}^\dagger r^i r^j
r^k \, g D^iD^j E^k \,{\rm S} \right\}$. However, the contribution of this operator vanishes 
for reasons similar to the ones that led to the vanishing of the 
leading-order thermal contribution in Eqs.~(\ref{leadingT11}) and (\ref{leadingT21}).
Second, they may involve diagrams with three or more insertions of the 
operators ${\rm Tr} \left\{  {\rm O}^\dagger {\vec r} \cdot g{\vec E} \,{\rm S} \right\}$ 
or ${\rm Tr} \left\{  {\rm O}^\dagger {\vec r} \cdot g{\vec E} \, {\rm O} \right\}$.
They contribute to order  $g^2\, r^2\, T^3 \times g^2 \, rT$ or higher to the thermal part 
of the static potential and hence are suppressed by at least a factor $rT$ with respect to the 
corrections considered in the previous paragraph. We will neglect them in the following.

\subsubsection{Contributions from the scale $m_D$}
\label{secshortmD}
In the previous section, having integrated out $T$ has led to a new EFT, which in the Yang--Mills
sector coincides with the HTL EFT and in the singlet sector shows a potential that 
is the sum of the terms (\ref{Vssoft}), (\ref{VsT}) and (\ref{VsTloop}).
In the weak-coupling regime, the Debye mass $m_D$, which is given by Eq.~(\ref{mD}), 
is smaller than the temperature. We assume that $m_D$ is the most relevant scale in the new EFT. 
The leading contributions to the static potential from the scale $m_D$ 
originate from the diagram shown in Fig.~\ref{figEE} when integrated over momenta $k \sim m_D$.
They are given by 
\bea
\left[ \delta {\bf V_s}(r)\right]_{11} &=& 
- i g^2 \, C_F \, \frac{r^2}{d-1}
\mu^{4-d} \int \frac{d^dk}{(2\pi)^d}
\frac{i}{-k^0 +i\epsilon}
\left[(k^0)^2 \, {\bf D}_{ii}(k) + \vec k^2 \, {\bf D}_{00}(k) 
\right]_{11}
\nn\\
&=&
- i g^2 \, C_F \, \frac{r^2}{d-1}
\mu^{4-d} \int \frac{d^dk}{(2\pi)^d} \, \pi \delta(-k^0) \, 
\left[(k^0)^2 \, {\bf D}_{ii}(k) + \vec k^2 \, {\bf D}_{00}(k) 
\right]_{11}\,,
\label{EETmD11}
\\
\left[ \delta {\bf V_s}(r)\right]_{22} &=& - \left[ \delta {\bf V_s}\right]^*_{11} \,,
\label{EETmD22}
\\
\left[ \delta {\bf V_s}(r)\right]_{12} &=& 0\,,
\label{EETmD12}
\\
\left[ \delta {\bf V_s}(r)\right]_{21} &=& 
i g^2 \, C_F \, \frac{r^2}{d-1}
\mu^{4-d} \int \frac{d^dk}{(2\pi)^d}
\, 2\pi \delta(-k^0) \, 
\left[(k^0)^2 \, {\bf D}_{ii}(k) + \vec k^2 \, {\bf D}_{00}(k) 
\right]_{21}\,,
\label{EETmD21}
\eea
where ${\bf D}_{\mu\nu}(k)$ is now the HTL resummed propagator. After integration in $k^0$
only ${\bf D}_{00}(0,\vec k)$ contributes, the expression of which can be found in (\ref{D00HTLresumk0}).
Substituting and performing the dimensional integrals, we obtain 
\bea
\delta {\bf V_s}(r)
&=&  \frac{C_F}{6} \, \als \, r^2m_D^3 \left(
\begin{matrix}
&&\hspace{-2mm}1
&&0 \\ 
&&\hspace{-2mm}0
&&-1
\end{matrix}
\right)
\nn\\
&&
-i \frac{C_F}{6} \, \als \, r^2 \, T \, m_D^2
\left(\frac{1}{\epsilon}-\gamma_E + \ln \pi + \ln \frac{\mu^2}{m_D^2} + \frac{5}{3} \right)
\left(
\begin{matrix}
&&\hspace{-2mm}1
&&0 \\ 
&&\hspace{-2mm}-2
&&1
\end{matrix}
\right)\,.
\label{VsmD}
\eea
Equation (\ref{VsmD}) shows that the scale $m_D$ contributes at order 
$g^2\, r^2\, m_D^3$ to the real part of the potential and at 
order  $g^2\, r^2\, T \, m_D^2$ to the imaginary one.

The Debye mass effectively plays the role of a gluon mass; in this sense, 
the real part of (\ref{VsmD}) agrees with a result that can be found in \cite{Brambilla:1999xf}.
The imaginary part originates, as the one in Eq.~(\ref{VsTloop}), 
from the imaginary part of the gluon self energy.  
It shows an ultraviolet divergence. This cancels against 
the infrared divergence of Eq.~(\ref{VsTloop}), which, we recall, comes 
from the diagram in Fig.~\ref{figEEmD} when integrated over momenta of order $T$.
We have already commented on this cancellation at the end of the previous section, 
we will add further specifications in Sec.~\ref{commentI}.

\subsection{Singlet static energy for $T\siml \Delta V$}
\label{secstaticenergy}
If $T \sim \Delta V$ there are no temperature dependent corrections to the potential, but
Eq.~(\ref{dVT11}) gives the leading thermal correction to the pole of the color singlet propagator. 
The real part of the pole provides the static energy of a color 
singlet quark-antiquark pair, whose leading thermal part is 
\be
\delta E_s = 
\frac{2}{3}\, N_c C_F \, \frac{\als^2}{\pi} \, r \, T^2\, f\!\left({N_c\als}/{(2rT)} \right)\,,
\label{EsdVT11}
\ee
where the function $f$ has been defined in (\ref{ffunction}).
Minus twice the imaginary part of the pole provides the color singlet thermal decay width,  
whose leading contribution, due to the decay of a static quark-antiquark 
color singlet into a quark-antiquark color octet, is 
\be
\Gamma 
= \frac{N_c^3C_F}{6} \, \frac{\als^4}{r}\, n_{\rm B}\!\left({N_c\als}/{(2r)}\right)\,.
\label{width}
\ee
In the situation $T \ll \Delta V$, the thermal width is exponentially suppressed, while 
the thermal contribution to the static energy becomes
\be
\delta E_s = - \frac{8}{45}\, \pi^3 \, \frac{C_F}{N_c} \, r^3\, T^4 
= - \frac{4}{3}\, \pi\, \frac{C_F}{N_c} \, r^3\,  
\left.\langle \vec E^a(0)\cdot \vec E^a(0)\rangle_T\right|_{\hbox{thermal part}}\,.
\label{EsdVT11condensate}
\ee
Equation (\ref{EsdVT11condensate}) agrees  with the analogous expression for the leading gluon 
condensate correction to the quark-antiquark static energy at zero temperature 
that was derived in \cite{Flory:1982qx}.

\subsection{Summary and comments}
\label{commentI}
As in the zero temperature case also in a thermal bath, the computation of  
the real-time potential between a static quark-antiquark pair requires integrating out 
all modes of energy and momentum larger than $\Delta V$. Modes of energy or momentum of order $\Delta V$ 
or smaller enter in physical observables, but, since they depend on the binding energy, 
do not belong to a proper EFT definition of the potential \cite{Brambilla:2004jw}.

If $ 1/r \gg T \gg m_D \gg \Delta V$ and in the weak-coupling regime, 
the static potential of a quark-antiquark pair 
is obtained by adding Eqs.~(\ref{Vssoft}), (\ref{VsT}), (\ref{VsTloop}) and (\ref{VsmD}).
In its diagonal form, it is given by 
\be
{\bf V_s}(r)
= [{\bf U}^{(0)}]^{-1}
\left(
\begin{matrix}
&&\hspace{-2mm}V_s(r)
&&0 \\ 
&&\hspace{-2mm}0
&&-V_s(r)^*
\end{matrix}
\right)
[{\bf U}^{(0)}]^{-1}\,,
\label{sumV11diag}
\ee
where 
\bea
V_s(r) &=& -C_F \frac{\alVs(1/r)}{r}
\nn\\
& & 
\hspace{-12mm} 
+ \frac{\pi}{9} \, N_c C_F \, \als^2 \, r \, T^2\, 
- \frac{3}{2} \zeta(3)\,  C_F \, \frac{\als}{\pi} \, r^2 \, T \,m_D^2
+ \frac{2}{3} \zeta(3)\, N_c C_F \, \als^2 \, r^2 \, T^3 
+ \frac{C_F}{6} \, \als \, r^2m_D^3  + \dots 
\nn\\
& &
\hspace{-12mm}
+ i \left[ 
- \frac{N_c^2 C_F}{6} \, \als^3\, T\,  
+ \frac{C_F}{6} \als \, r^2 \, T \,m_D^2\, \left( 2 \gamma_E 
- \ln\frac{T^2}{m_D^2} -1 - 4 \ln 2 - 2 \frac{\zeta^\prime(2)}{\zeta(2)} \right)
\right.
\nn\\
&& \quad \left.
+ \frac{4\pi}{9} \ln 2 \; N_c C_F \,  \als^2\, r^2 \, T^3
\right]
+ \dots 
\,.
\label{sumV11}
\eea
The leading term is provided by the Coulomb part: $ -C_F \als/r$. 
In the real part, the first thermal correction is of order 
$g^2 \, r^2 \, T^3 \times \Delta V/T$, the second and third ones are of order 
$g^2 \, r^2 \, T^3 \times (m_D/T)^2$ and the last one is of order $g^2\, r^2\, m_D^3$.
The correction proportional to $\als\,r^2\,m_D^3$ is suppressed by
$m_D/T$ with respect to the one proportional to $\als\, r^2\,T\, m_D^2$ and 
may be neglected. In the imaginary part, the first term is of order  
$g^2 \, r^2 \, T^3 \times (\Delta V/T)^2$ and the other ones are of order $g^2 \, r^2 \, T^3 \times (m_D/T)^2$.
The dots in (\ref{sumV11}) stand for higher-order real and imaginary terms. 
Temperature dependent higher-order terms are suppressed 
by powers of $\Delta V/T$, $m_D/T$, $rT$, $\Delta V/m_D$ and $rm_D$.

The imaginary part of $V_s(r)$ has two origins. The first term comes from the thermal break up 
of a quark-antiquark color singlet state into a color octet state. The other terms come from imaginary 
contributions to the gluon self energy that may be traced back to the Landau damping phenomenon.
Both are thermal effects: the first one is specific of the non-Abelian nature of QCD, 
the second one would also show up in QED, although in QED the photon polarization tensor would get 
only fermionic contributions. Having assumed $ T \gg m_D \gg \Delta V$, 
the term due to the singlet to octet break up is suppressed by $(\Delta V/m_D)^2$ with respect 
to the imaginary gluon self-energy contributions, which provide therefore the 
parametrically leading contribution to the imaginary part of the potential.

The thermal part of Eq.~(\ref{sumV11}) is finite because, under 
the condition $ 1/r \gg T \gg m_D \gg \Delta V$, it provides 
the leading thermal contribution to the real-time energy and to the decay width 
of a static quark and antiquark pair  in a color-singlet configuration. 
Note, however, that the non-thermal part of the potential, $\alVs$, 
develops infrared divergences starting from order $\als^4$, which eventually cancel 
in physical quantities against contributions from the scale $\Delta V$.
These have been most recently considered in \cite{Brambilla:2006wp}. 
If $1/r \gg T \gg \Delta V \simg m_D$, the static potential is the sum of 
Eqs.~(\ref{Vssoft}), (\ref{VsT}) and (\ref{VsTloop}). The thermal part is infrared divergent.
Divergences cancel in physical observables against thermal contributions coming 
from the scale $m_D$. Finally, if $1/r \gg \Delta V \simg T$, the static potential is just its Coulomb part 
(\ref{Vssoft}). Thermal contributions affect physical observables through loop corrections 
that involve momenta or energies of the order of the binding energy or smaller, but do not 
modify the potential. In this case, the leading thermal effect on the static energy of the 
color singlet quark-antiquark pair can be read in (\ref{EsdVT11}) and the leading thermal 
width, which is due to the singlet to octet break-up phenomenon, in (\ref{width}).

Our result is also relevant for the case of a quark and antiquark 
with a large but finite mass $m$. This would correspond to the actual case of 
heavy quarkonium in a thermal medium. However, in the case of finite mass, the relevant scales 
of the bound state are dynamical and the above discussion gets modified accordingly.
For a comprehensive review in the $T=0$ case we refer to \cite{Brambilla:2004jw}. 
Bound states of heavy quarks are characterized by the energy scales $m$, $mv$, $mv^2$, 
where $v$ is the relative velocity of the heavy quarks: $mv$ is of the order of the inverse size of the 
bound state and $mv^2$ is of the order of the kinetic or binding energy.
The equation of motion of the quark-antiquark bound state reduces in the non-relativistic 
limit to the Schr\"odinger equation. The potential is the interaction term entering 
the Schr\"odinger equation. Only modes of energy and momentum larger than $mv^2$ contribute 
to the potential.  With these specifications, Eq.~(\ref{sumV11}) provides the static potential of 
a heavy quarkonium in a thermal medium under the condition $ m \gg mv \gg T \gg m_D \gg mv^2$. 
If $ m \gg mv \gg T \gg mv^2 \simg m_D$, the heavy quarkonium static potential 
is given by the sum of Eqs.~(\ref{Vssoft}) (\ref{VsT}) and (\ref{VsTloop}) only. 
The potential turns out to be infrared divergent also in its thermal 
part and the comment at the end of paragraph {\bf (2)} of Sec.~\ref{secshortT} applies. 
The divergence will eventually cancel in physical quantities against contributions 
coming from the scale $mv^2$ or lower. If $ m \gg mv \gg mv^2 \simg T \gg  m_D$, 
the heavy quarkonium static potential is just the Coulomb potential (\ref{Vssoft}) and 
thermal corrections enter physical quantities in loop involving momenta at the scale 
$mv^2$ or lower.

In the next section, we provide a derivation of Eq.~(\ref{sumV11}) that does 
not make use of the EFT language and follows directly from a calculation of the potential 
in perturbative QCD.

\section{Short-distance thermal corrections to the potential in perturbative QCD}
\label{secpQCD}
\begin{figure}[ht]
\makebox[-5truecm]{\phantom b}
\put(0,0){\epsfxsize=4truecm \epsfbox{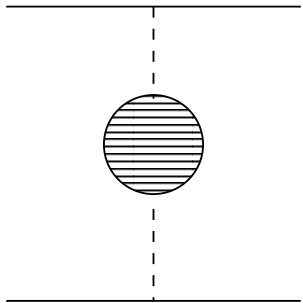}}
\caption{Longitudinal gluon exchange between a static quark and a static antiquark; 
the dashed blob stands for the gluon self energy.}  
\label{figgluon00}
\end{figure}

\begin{figure}[ht]
\makebox[-5truecm]{\phantom b}
\put(0,0){\epsfxsize=5truecm \epsfbox{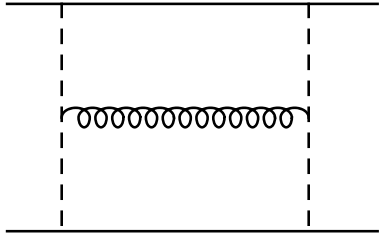}}
\caption{Box diagram: the upper line represents a static quark and the lower one a static antiquark.}  
\label{figuswilson0}
\end{figure}

In this section, we ask the question of what would be the origin of 
the thermal part of the potential given by Eq.~(\ref{sumV11})
if we would not introduce any EFT treatment, but simply perform a calculation in 
perturbative QCD under the condition that $1/r \gg T$ and that the binding energy
is much smaller than $m_D$. The answer is that the thermal part of (\ref{sumV11}) would originate 
from the longitudinal gluon exchange, with a self-energy insertion, between 
a static quark and a static antiquark shown in Fig.~\ref{figgluon00} and 
from the box diagram shown in Fig.~\ref{figuswilson0}.

We first consider the diagram in Fig.~\ref{figgluon00}, which 
contributes to the physical ``11'' component of the static potential by 
\be
\left[ \delta {\bf V_s}(r)\right]_{11} = 
\mu^{4-d} \int \frac{d^{d-1}k}{(2\pi)^{d-1}} \, e^{-i \vec k \cdot \vec r}
\,g^2 \, C_F \, \left[ i \delta {\bf D}_{00}(0,\vec k)\right]_{11}\,,
\label{VsQCD}
\ee
where $\delta {\bf D}_{00}(k)$ is defined in Eqs.~(\ref{D0011})-(\ref{DRAPi}) and 
depends on the gluon polarization $\Pi_{00}^{\rm R,A}$.
Note that we have set to zero the fourth-component of the momentum in the longitudinal gluon: 
corrections would be suppressed by powers of $k^0/|\vec k| \sim V_s\,r$ or $V_s/T$ or $V_s/m_D$.
Equation (\ref{VsQCD}) gets contributions from different momentum regions. 

{\bf (1)} The first  momentum region is $|\vec k| \sim 1/r$. The thermal contribution to the 
longitudinal gluon polarization tensor when $|\vec k| \sim 1/r \gg T$ 
is provided by Eq.~(\ref{Pi00koo}), which, substituted in 
Eq.~(\ref{VsQCD}), gives (the integral is finite, hence $d=4$)
\be
\left[ \delta {\bf V_s}(r)\right]_{11} = 
\int \frac{d^3k}{(2\pi)^3}e^{-i\vec k \cdot \vec r} 
\left(-C_F \frac{4\pi\als}{\vec k^4}\right) \frac{N_c\,g^2\,T^2}{18}
= \frac{\pi}{9} \, N_c C_F \, \als^2 \, r \, T^2\,,
\label{VsTbis}
\ee
where we have used that the Fourier transform of $4\pi/\vec k^4$ is $-r/2$.
Equation (\ref{VsTbis}) agrees with the real part of Eq.~(\ref{VsT}).

\begin{figure}[ht]
\makebox[-7truecm]{\phantom b}
\put(0,0){\epsfxsize=7truecm \epsfbox{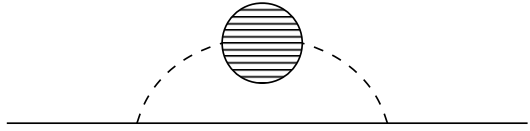}}
\caption{Gluon self-energy correction to the one-loop self-energy diagram of a static quark.}  
\label{figself00}
\end{figure}

{\bf (2)} A second momentum region is $|\vec k|\sim T$. Since $T \ll 1/r$, under the condition 
$|\vec k| \sim T$ we may expand the exponential 
$\displaystyle e^{-i \vec k \cdot \vec r}$ in (\ref{VsQCD}):
\be
\left[ \delta {\bf V_s}(r)\right]_{11} = 
\mu^{4-d} \int \frac{d^{d-1}k}{(2\pi)^{d-1}} \, \left(1 - \frac{(\vec k \cdot \vec r)^2}{2} + \dots \right) 
\,g^2 \, C_F \, \left[ i \delta {\bf D}_{00}(0,\vec k)\right]_{11}\,.
\label{VsQCDmultipole}
\ee
The first term in the expansion corresponds to a mass correction and cancels against twice the 
thermal contribution of the static quark self energy with a gluon self-energy insertion, 
see Fig.~\ref{figself00}. The second term coincides with the expression in Eq.~(\ref{Vs11Teq1}) 
and gives the same result as (\ref{VsTloop}).

{\bf (3)} Finally, a third momentum region is  $|\vec k|\sim m_D$. The contribution to the potential is like 
Eq.~(\ref{VsQCDmultipole}), but now $|\vec k|\sim m_D \ll T$ and the correct expression 
for ${\bf D}_{00}(0,\vec k)$ is the HTL resummed propagator given in (\ref{D00HTLresumk0}).
The first term in the expansion corresponds to a mass correction, which this time comes from the scale 
$m_D$ and cancels against twice the contribution of the static quark self energy, 
see Fig.~\ref{figself}, when the loop momentum is of order $m_D$ and a HTL resummed gluon 
propagator is used. The second term gives the same result as (\ref{VsmD}).

We consider now the diagram of Fig.~\ref{figuswilson0}, which 
contributes to the physical ``11'' component of the static potential by 
(we write the thermal part only)
\bea
\left[ \delta {\bf V_s}(r)\right]_{11} &=& 
\int \frac{d^3k}{(2\pi)^{3}} \, e^{-i \vec k \cdot \vec r}
\left[i\,N_c^2C_F\,g^6\int \frac{d^4p}{(2\pi)^{4}}\int \frac{d^4q}{(2\pi)^{4}}
\,p^i(k^j-p^j)\,\frac{i}{-p^0+i\epsilon}\frac{i}{p^0-q^0+i\epsilon} \right.
\nn\\
&& \times \left. \left(\delta_{ij} - \frac{q^iq^j}{\vec q^2}\right)
\, 2\pi\delta(q^2) \,n_{\rm B}(|q^0|) \, 
\frac{i}{\vec p^2}\frac{i}{|\vec p - \vec q|^2}\frac{i}{|\vec k - \vec p|^2}
\frac{i}{|\vec k - \vec p + \vec q|^2} \right]\,.
\label{VsTtris}
\eea
The imaginary part of the integral comes from the real part 
\bea
{\rm Re}\, \int \frac{dp^0}{2\pi} \frac{i}{-p^0+i\epsilon}\frac{i}{p^0-q^0+i\epsilon}
= {\rm Re}\, \frac{i}{-q^0+i\epsilon} = \pi \, \delta(-q^0)\,,
\nn
\eea
which inserted in (\ref{VsTtris}) gives the imaginary part of (\ref{VsT}). The result 
is exact and does not rely on any expansion in the kinematical variables.

The sum of the contributions coming from the three momentum regions in the integral (\ref{VsQCD})
and from the imaginary part of (\ref{VsTtris}) gives Eq.~(\ref{sumV11}).
We note that also the calculation in perturbative QCD shows that the imaginary part 
of Eq.~(\ref{VsT}) has a different origin from the other ones:
it comes from the box diagram of Fig.~\ref{figuswilson0}, which describes  
a singlet to octet to singlet transition, while the other ones come 
from the gluon self-energy diagram of Fig.~\ref{figgluon00}.

\section{Bound states for $1/r\ll T$}
\label{secSpNRQCD}
In this section, we consider bound states made of a static quark and antiquark in a thermal 
bath at distances such that $1/r\ll T$. We still keep that $T$, $1/r$ and $m_D$ are 
perturbative scales. We further neglect other thermodynamical scales.

Under the above condition, the first scale to integrate out from QCD is the temperature $T$.
At one loop, this was done in Sec.~\ref{secHTL}. Integrating out $T$ 
leads in the Yang--Mills sector to the HTL Lagrangian \cite{Braaten:1991gm}.
In the heavy-quark sector, one loop contributions vanish in the static limit. 
At two loop, see for instance the diagram in Fig.~\ref{figself00}, there may be effects. 
These are of order $\als^2\, T$ $\sim$ $\als\,T\,(m_D/T)^2$ $\sim$ $ \als\, m_D \, (m_D/T)$ 
and will be neglected in the following where we shall concentrate 
on the leading contribution coming from the scale $m_D$.

The next scale to integrate out is $1/r$. We assume $1/r \simg m_D$, and integrate out 
both scales $1/r$ and $m_D$ at the same time. We shall specialize to the case 
$1/r \gg m_D$ in Sec.~\ref{secTrmD}.

\subsection{Singlet and octet propagators}
\label{secVIA}
After integrating out the scales $1/r \simg m_D$ it may be convenient to introduce 
quark-antiquark fields in analogy with what was done in previous sections.
The real-time quark-antiquark propagator, ${\bf S}(p)$, is  
a $2 \times 2$ matrix obtained by matching equal time quark and antiquark propagators 
such that $[{\bf S}(p)]_{ij}$ provides the propagator of a quark-antiquark pair of type 
``$i$'' into a quark-antiquark pair of type ``$j$''. The explicit expressions of the 
free color singlet and color octet quark-antiquark propagators are like those derived in 
Sec.~\ref{secsingletoctetrT}:
\bea
{\bf S}^{\rm singlet\, (0)}(p) &=& 
\left(
\begin{matrix}
&&\hspace{-2mm}\displaystyle\frac{i}{p^0 +i\epsilon}  
&&0 \\ 
&&\hspace{-2mm}2\pi\delta(p^0)  
&&\displaystyle\frac{-i}{p^0 -i\epsilon}
\end{matrix}
\right)
= 
{\bf U}^{(0)} 
\left(
\begin{matrix}
&&\hspace{-2mm}\displaystyle\frac{i}{p^0 +i\epsilon} 
&&0 \\ 
&&0  
&&\displaystyle\frac{-i}{p^0 -i\epsilon}
\end{matrix}
\right)
{\bf U}^{(0)}\,,
\label{SsingletstaticrT}
\\
{\bf S}^{\rm octet\, (0)}(p)_{ab} &=& 
\delta_{ab}\,\left(
\begin{matrix}
&&\hspace{-2mm}\displaystyle\frac{i}{p^0 +i\epsilon}
&&0 \\ 
&&\hspace{-2mm}2\pi\delta(p^0)
&&\displaystyle\frac{-i}{p^0 - i\epsilon}
\end{matrix}
\right)
= 
\delta_{ab}\,{\bf U}^{(0)} 
\left(
\begin{matrix}
&&\hspace{-2mm}\displaystyle\frac{i}{p^0 +i\epsilon}
&&0 \\ 
&&0  
&&\displaystyle\frac{-i}{p^0-i\epsilon}
\end{matrix}
\right)
{\bf U}^{(0)}\,.
\label{SoctetstaticrT}
\eea
Singlet and octet fields have been normalized as in Sec.~\ref{secWpNRQCD}.

Thermal contributions from the scales $1/r \simg m_D$ modify the quark-antiquark propagator.
In particular, the singlet propagator gets the form 
\bea
{\bf S}^{\rm singlet}(p) &=& 
\left(
\begin{matrix}
&&\hspace{-2mm}\displaystyle\frac{i}{p^0 - \delta m -V_s(r) +i\epsilon}  
&&0 \\ 
&&\hspace{-2mm} \displaystyle  
\frac{i}{p^0 - \delta m -V_s(r) +i\epsilon}  - \frac{i}{p^0 - \delta m^* -V_s^*(r) -i\epsilon}  
&&\displaystyle\frac{-i}{p^0  - \delta m^* -V_s^*(r) -i\epsilon}
\end{matrix}
\right)
\nn\\
\nn\\
&=& 
{\bf S}^{\rm singlet\, (0)}(p) + 
{\bf S}^{\rm singlet\, (0)}(p) \left[ -i\,\delta {\bf m} -i\, {\bf V_s} \right] {\bf S}^{\rm singlet\, (0)}(p) 
+ \dots \; ,
\label{Ssingletstaticfull}
\eea
where in the last line we have expanded with respect to $\delta m$ and $V_s$ and introduced the $2\times 2$ 
matrices:
\bea
\delta {\bf m}
&=& 
\left(
\begin{matrix}
&&\hspace{-2mm}\delta m
&&0 \\ 
&&\hspace{-2mm} -2i \, {\rm Im} \, \delta m
&&\displaystyle -\delta m^*
\end{matrix}
\right)
=
[{\bf U}^{(0)}]^{-1}
\left(
\begin{matrix}
&&\hspace{-2mm} \delta m
&&0 \\ 
&&\hspace{-2mm}0
&&- \delta m^*
\end{matrix}
\right)
[{\bf U}^{(0)}]^{-1}\,,
\label{diagonalm}
\\
{\bf V_s}
&=& 
\left(
\begin{matrix}
&&\hspace{-2mm} V_s
&&0 \\ 
&&\hspace{-2mm} -2i \, {\rm Im} \, V_s
&&\displaystyle - V_s^*
\end{matrix}
\right)
=
[{\bf U}^{(0)}]^{-1}
\left(
\begin{matrix}
&&\hspace{-2mm} V_s
&&0 \\ 
&&\hspace{-2mm}0
&&- V_s^*
\end{matrix}
\right)
[{\bf U}^{(0)}]^{-1}\,.
\label{diagonalVs}
\eea

\subsection{Matching the mass term $\delta m$}
\label{secVIB}
The static quark (antiquark) self energy at one loop is shown in Fig.~\ref{figself}.
In the case considered here, the loop momentum is of order $m_D$ and the HTL resummed 
gluon propagator is used. We match in the real-time formalism the self energy diagram 
(normalized in color space) with the second term in the expansion (\ref{Ssingletstaticfull}), 
${\bf S}^{\rm singlet\, (0)}(p) \left[-i\, \delta {\bf m}\right]  {\bf S}^{\rm singlet\, (0)}(p)$,  
obtaining:
\bea
[\delta {\bf m}]_{11} &=& 
i\, (ig)^2\,C_F\, \mu^{4-d}\int \frac{d^dk}{(2\pi)^d} 
\left[ \frac{i}{-k^0+i\epsilon} - \frac{i}{-k^0-i\epsilon} \right] \, [{\bf D}_{00}(k)]_{11} 
\nn\\
&=&
i\, (ig)^2 \,C_F\,\mu^{4-d}\int \frac{d^dk}{(2\pi)^d} \, 2\pi \delta(-k^0)\,[{\bf D}_{00}(k)]_{11} 
= - C_F\, \als \left( m_D + i T \right)\,,
\label{deltam11}\\ 
{[\delta {\bf m}]}_{22} &=& - [\delta {\bf m}]^*_{11}\,,
\label{deltam22}\\ 
{[\delta {\bf m}]}_{12} &=& 0\,,
\label{deltam12}\\ 
{[\delta {\bf m}]}_{21} &=& 
i\, g^2 \,C_F\,\mu^{4-d}\int \frac{d^{d}k}{(2\pi)^{d}} \,  
\left[ 2\pi \delta(-k^0) + 2\pi \delta(-k^0) \right] \,[{\bf D}_{00}(k)]_{21}
= 2i \, C_F \, \als\, T \,,
\label{deltam21}
\eea
where $i/(-k^0+i\epsilon)$ and $ - i/(-k^0-i\epsilon)$ are the ``11'' components 
of the static quark and antiquark propagators respectively, $2\pi \delta(-k^0)$ is the 
``21'' component of both static quark and antiquark propagators and 
we have added the contributions from the quark and the antiquark; 
equation (\ref{deltam12}) vanishes because the component 
``12'' of the heavy quark and antiquark propagators vanishes, see Eqs.~(\ref{Squarkstatic}) 
and (\ref{Santiquarkstatic}). ${\bf D}_{00}(k)$ is the longitudinal HTL resummed gluon propagator, 
whose expression for $k^0=0$ is given in Eq.~(\ref{D00HTLresumk0}). 

The matrix $\delta {\bf m}$ has indeed the form (\ref{diagonalm}); it is diagonalized 
by the matrix $ \displaystyle[{\bf U}^{(0)}]^{-1}$ and $\delta m$ is given by 
\be
\delta m = - C_F\, \als \left( m_D + i T \right)\,.
\label{deltam}
\ee
The real part of  $\delta m$ corresponds to the free energy of two isolated 
static quarks in the imaginary-time formalism, which was first calculated 
in Ref.~\cite{Gava:1981qd} (for further discussions see Ref.~\cite{Petreczky:2005bd}).
The imaginary part of $\delta m$ is minus twice the damping rate 
of an infinitely heavy fermion \cite{Pisarski:1993rf}.

\subsection{Matching the singlet static potential $V_s$}
\label{secVIC}
\begin{figure}[ht]
\makebox[-12truecm]{\phantom b}
\put(0,0){\epsfxsize=3truecm \epsfbox{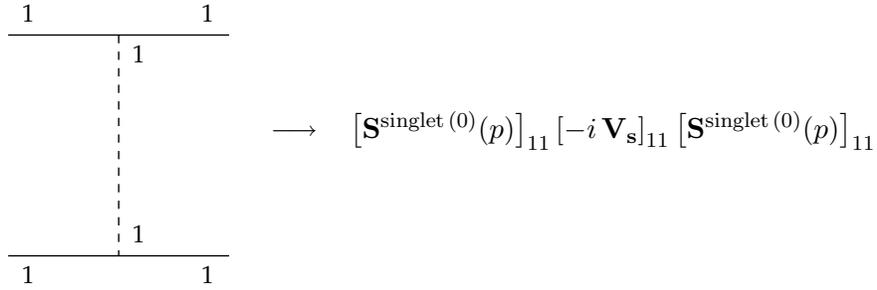}}
\put(130,55){$\left[{\bf S}^{\rm singlet\, (0)}(p)\right]_{11} \left[-i\, {\bf V_s}\right]_{11}  
\left[{\bf S}^{\rm singlet\, (0)}(p)\right]_{11}$}
\put(100,55){$\longrightarrow$}
\caption {Matching condition for $\left[-i\, {\bf V_s}\right]_{11}$; the numbers label the type ``1'' and 
``2'' propagators. All entries in a vertex are of the same type. Vertices of type ``1'' and ``2'' have opposite 
signs.}
\label{figD11}
\end{figure}

\begin{figure}[ht]
\makebox[-12truecm]{\phantom b}
\put(0,0){\epsfxsize=3truecm \epsfbox{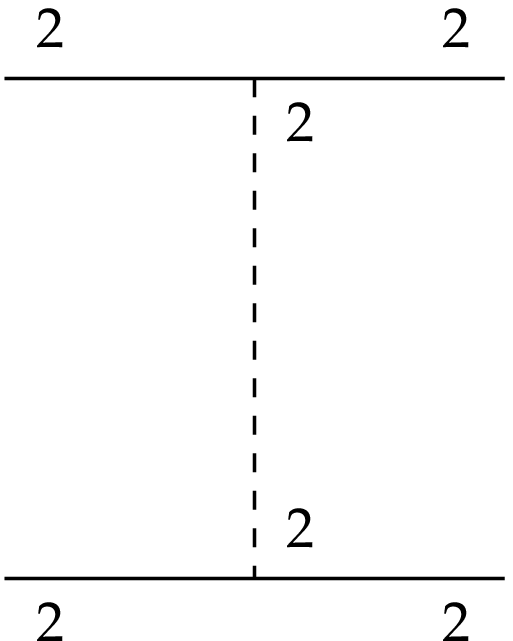}}
\put(130,55){$\left[{\bf S}^{\rm singlet\, (0)}(p)\right]_{22} \left[-i\, {\bf V_s}\right]_{22}  
\left[{\bf S}^{\rm singlet\, (0)}(p)\right]_{22}$}
\put(100,55){$\longrightarrow$}
\caption {Matching condition for $\left[-i\, {\bf V_s}\right]_{22}$.}
\label{figD22}
\end{figure}

\begin{figure}[ht]
\makebox[-16truecm]{\phantom b}
\put(0,0){\epsfxsize=7.5truecm \epsfbox{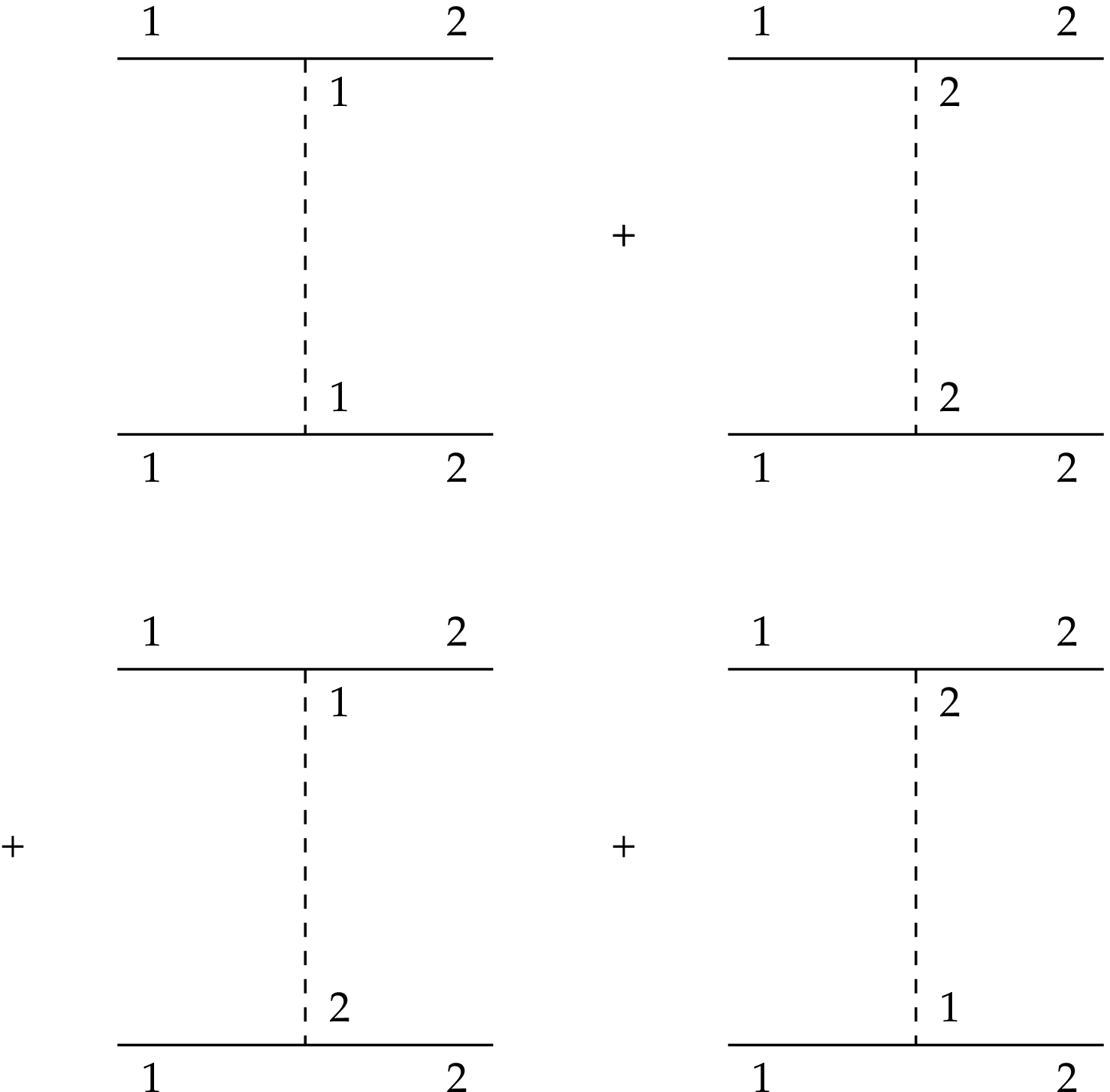}}
\put(240,100){
$\left[{\bf S}^{\rm singlet\, (0)}(p)\right]_{1i} \left[-i\, {\bf V_s}\right]_{ij}  
\left[{\bf S}^{\rm singlet\, (0)}(p)\right]_{j2}$ 
}
\put(220,100){$\longrightarrow$}
\caption {Matching condition for $\left[-i\, {\bf V_s}\right]_{12}$.}
\label{figD12}
\end{figure}

\begin{figure}[ht]
\makebox[-16truecm]{\phantom b}
\put(0,0){\epsfxsize=7.5truecm \epsfbox{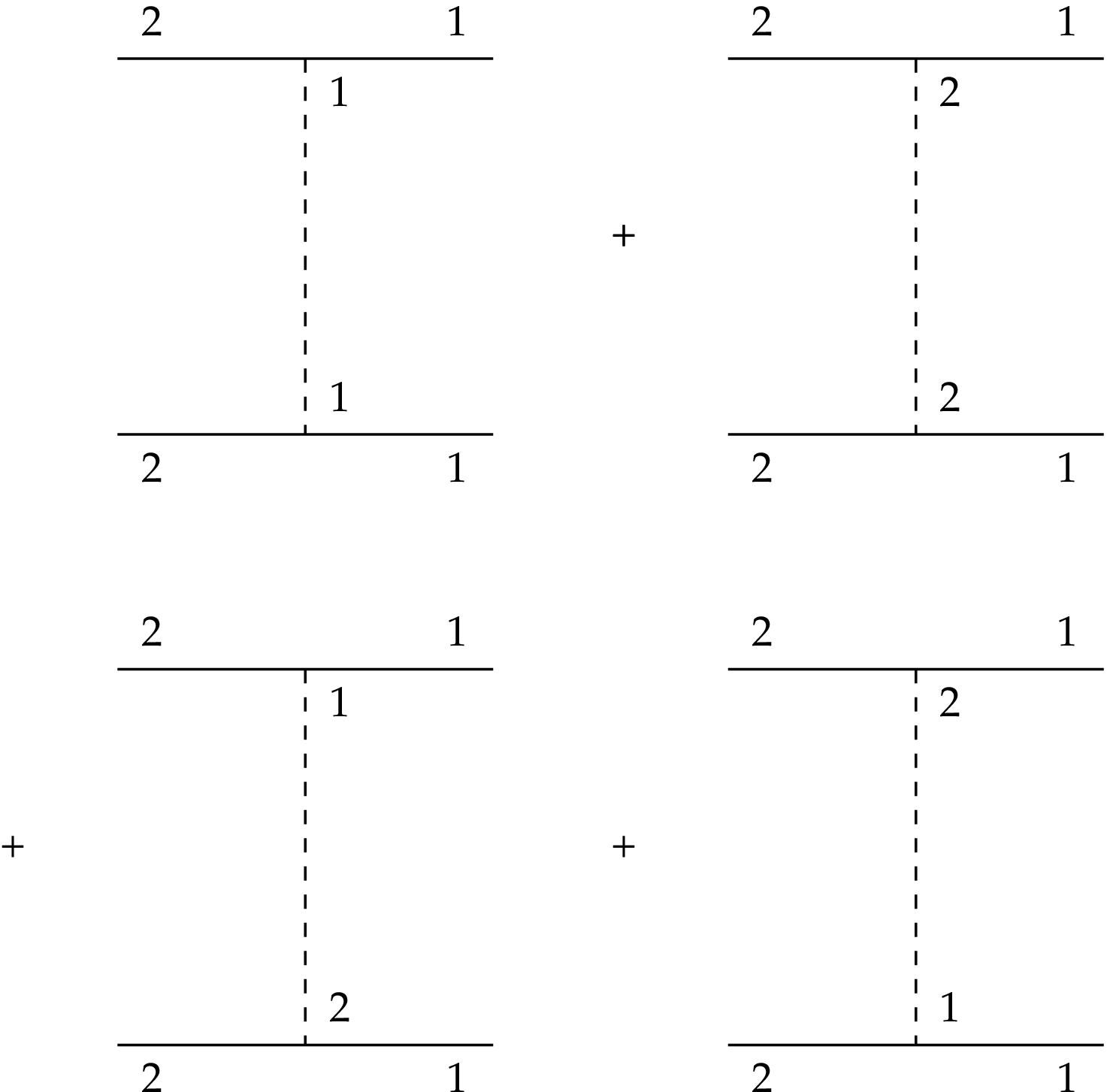}}
\put(240,100){
$\left[{\bf S}^{\rm singlet\, (0)}(p)\right]_{2i} \left[-i\, {\bf V_s}\right]_{ij}  
\left[{\bf S}^{\rm singlet\, (0)}(p)\right]_{j1}$
}
\put(220,100){$\longrightarrow$}
\caption {Matching condition for $\left[-i\, {\bf V_s}\right]_{21}$.}
\label{figD21}
\end{figure}

The matrix elements $[{\bf V_s}]_{ij}$ are obtained by matching in real time one-gluon 
exchange diagrams that transform a quark-antiquark pair of type ``$i$'' into a quark-antiquark pair 
of type ``$j$'' with the third term in the expansion (\ref{Ssingletstaticfull}), 
${\bf S}^{\rm singlet\, (0)}(p)$ $\left[-i\, {\bf V_s}\right]$  ${\bf S}^{\rm singlet\, (0)}(p)$.
More precisely, by matching the diagram of Fig.~\ref{figD11} with 
$\left[{\bf S}^{\rm singlet\, (0)}(p)\right]_{11}$ $\left[-i\, {\bf V_s}\right]_{11}  
\left[{\bf S}^{\rm singlet\, (0)}(p)\right]_{11}$, the diagram of Fig.~\ref{figD22} with 
$\left[{\bf S}^{\rm singlet\, (0)}(p)\right]_{22} \left[-i\, {\bf V_s}\right]_{22}  
\left[{\bf S}^{\rm singlet\, (0)}(p)\right]_{22}$, the diagrams of Fig.~\ref{figD12} with 
$\left[{\bf S}^{\rm singlet\, (0)}(p)\right]_{1i} \left[-i\, {\bf V_s}\right]_{ij}  
\left[{\bf S}^{\rm singlet\, (0)}(p)\right]_{j2}$ and the diagrams of   Fig.~\ref{figD21} with 
$\left[{\bf S}^{\rm singlet\, (0)}(p)\right]_{2i}$ $\left[-i\, {\bf V_s}\right]_{ij}$  
$\left[{\bf S}^{\rm singlet\, (0)}(p)\right]_{j1}$ 
we obtain  
\bea
[{\bf V_s}(r)]_{11} &=& 
\int \frac{d^3k}{(2\pi)^3}e^{-i\vec k \cdot \vec r}
\, i g^2\,C_F\, [{\bf D}_{00}(0,\vec k)]_{11} 
\nn\\
&=&
\int \frac{d^3k}{(2\pi)^3}e^{-i\vec k \cdot \vec r}
\left(
-C_F \frac{4\pi \als }{\vec k^2+m_D^2} 
+ i \, C_F \, \frac{T}{|\vec k|} \, m_D^2 \frac{4\pi^2\als }{(\vec k^2+m_D^2)^2}
\right)\,,
\label{Vs11}\\
{[{\bf V_s}(r)]}_{22} &=& - [{\bf V_s}(r)]^*_{11}\,, 
\label{Vs22}\\
{[{\bf V_s}(r)]}_{12} &=& 0\,,
\label{Vs12}\\
{[{\bf V_s}(r)]}_{21} &=& 
\int \frac{d^3k}{(2\pi)^3}e^{-i\vec k \cdot \vec r}
i (-g^2)\,C_F\, \left([{\bf D}_{00}(0,\vec k)]_{12} + [{\bf D}_{00}(0,\vec k)]_{21} \right)
\nn\\
&=& -2i\,{\rm Im} \, [{\bf V_s}(r)]_{11}\,,
\label{Vs21}
\eea
where the longitudinal HTL resummed gluon propagator, ${\bf D}_{00}(0,\vec k)$,  
given in Eq.~(\ref{D00HTLresumk0}), comes from expanding in the external energy, 
which is much smaller than the typical momentum $\sim 1/r$.

The matching conditions of Figs.~\ref{figD11} and \ref{figD22} fix 
$[{\bf V_s}(r)]_{11}$ and $[{\bf V_s}(r)]_{22}$ as in Eqs.~(\ref{Vs11}) and 
(\ref{Vs22}) respectively. In Figs.~\ref{figD12} and \ref{figD21} the first two diagrams 
cancel in the matching against the terms $i=j=1$ and $i=j=2$ in the sum on the right-hand side.
In the case of Fig.~\ref{figD12}, the last two diagrams vanish, because 
the ``12'' component of the static quark propagator vanishes, leading to 
Eq.~(\ref{Vs12}); while in the case of Fig.~\ref{figD21},  the last two diagrams 
give  $[{\bf V_s}(r)]_{21}$ as in Eq.~(\ref{Vs21}).

The matrix ${\bf V_s}$ has indeed the form (\ref{diagonalVs}); it is diagonalized 
by the matrix $ \displaystyle[{\bf U}^{(0)}]^{-1}$ and $V_s$ is given by 
\bea
V_s(r) &=& 
\int \frac{d^3k}{(2\pi)^3}e^{-i\vec k \cdot \vec r}
\left(-C_F \frac{4\pi \als }{\vec k^2+m_D^2} 
+ i \, C_F \, \frac{T}{|\vec k|} \, m_D^2 \frac{4\pi^2\als }{(\vec k^2+m_D^2)^2}
\right)
\nn\\
&=& -C_F\,\frac{\als}{r}\,e^{-m_Dr} 
+ iC_F\,\als\, T\,\frac{2}{rm_D}\int_0^\infty dx \,\frac{\sin(m_Dr\,x)}{(x^2+1)^2}
\,.
\label{Vs}
\eea
The expression of $V_s(r)$, which we have derived here in real-time formalism,
agrees with the analogous expression derived in imaginary-time formalism, 
after analytical continuation, in \cite{Laine:2006ns}. It should be emphasized 
that under the condition $1/r \sim m_D$ the real part of (\ref{Vs}) is of order 
$\als m_D$, hence subleading with respect to the imaginary part, which is 
of order $\als T$: the quark-antiquark pair decays before forming the bound state, 
whose typical time scale is proportional to the inverse of the real part of the potential.

Note that the short-distance expansion of Eq.~(\ref{Vs}) would give, up to order $r^0$, 
the Coulomb potential and an $r$-independent term, $C_F\, \als \left( m_D + i T \right)$, which 
would cancel the mass term derived in (\ref{deltam}). In the kinematical situation discussed here, 
this reflects the analogous cancellation between the mass correction and the potential 
correction induced by the scale $m_D$ discussed in paragraph {\bf (3)} of Sec.~\ref{secpQCD}.

In this section, we have assumed that the singlet propagator has the form
(\ref{Ssingletstaticfull}) and verified that this is indeed the case by performing the matching.
We could also have proceeded with the reverse logic used in the rest of the paper: 
by matching $\delta {\bf m}$ and $\delta {\bf V_s}$ we would have realized 
that these matrices are diagonalized by $[{\bf U}^{(0)}]^{-1}$, hence, 
that they fulfill Eq.~(\ref{sumSinglet}) and give rise to the resummed propagator (\ref{Ssingletstaticfull}).

\subsection{Singlet static energy for $1/r \sim m_D$}
Adding the real parts of Eqs.~(\ref{deltam}) and (\ref{Vs}) gives the leading static quark-antiquark energy 
for $1/r \sim m_D$: 
\be 
E_s = -C_F\,\als\,m_D -C_F\,\frac{\als}{r}\,e^{-m_Dr} \,,
\label{energy2}
\ee
and the imaginary parts of Eqs.~(\ref{deltam}) and (\ref{Vs}) provide the leading 
static quark-antiquark thermal decay width:
\be 
\Gamma = 2\,C_F\,\als\,T \left[
1 - \frac{2}{rm_D}\int_0^\infty dx \,\frac{\sin(m_Dr\,x)}{(x^2+1)^2} \right] \,.
\label{width2}
\ee
The thermal width originates from the imaginary part of the gluon self energy. Singlet to octet 
transitions contribute to the decay width as well and the leading contribution is provided 
by the diagram of Fig.~\ref{figuswilson0}, which gives (see the imaginary part of (\ref{VsT})):
$\displaystyle \delta\Gamma = \frac{1}{3} \, N_c^2 C_F \, \als^3\, T$. This contribution is parametrically  
suppressed by a factor $\als^2$, whose natural scale is of order $1/r$, with respect to the one in (\ref{width2}).
Note, however, that $\delta \Gamma$ may be numerically as large as 50\% of $\Gamma$ for $\als \approx 0.3$ 
and even larger than $\Gamma$ for $\als \approx 0.5$; see also Fig.~\ref{figRatioGamma}.

The static energy given by Eq. (\ref{energy2}) coincides with the leading-order result \cite{Petreczky:2005bd} 
of the so-called singlet free energy first introduced by Nadkarni \cite{Nadkarni:1986as} 
(the heavy quark-antiquark free energy was defined by McLerran and Svetitsky in \cite{McLerran:1981pb})
and also studied in lattice QCD (see e.g. \cite{Kaczmarek:2002mc,Kaczmarek:2004gv} 
and \cite{Brambilla:2004wf} for reviews). 
We recall that the free energy describes a thermodynamical property 
of the system and it is computed from the static quark-antiquark propagator evaluated at the imaginary 
time $1/T$ (for large temperatures this corresponds to small imaginary times), while 
the static energy studied in this work describes the real-time evolution of a quark-antiquark 
pair and it is computed by evaluating the quark-antiquark propagator at infinite real times.
The thermal decay width (\ref{width2}) coincides with the result of Ref.~\cite{Laine:2006ns}.

\subsection{The $1/r \gg m_D$ case}
\label{secTrmD} 
In the $1/r \gg m_D$ case, but with $m_D$ still larger than the binding energy, 
the scales $1/r$ and $m_D$ are integrated out in two subsequent matchings.
First, the matching at the scale $1/r$ can be done in close analogy with the discussion in sections
\ref{secVIA}-\ref{secVIC}. The potential is given by Eqs. (\ref{Vs11})-(\ref{Vs21}):  
since $|\vec k| \sim 1/r \gg m_D$ we expand ${\bf D}_{00}(0,\vec k)$ in powers of $m_D^2/\vec k^2$. 
We need to regularize the integrals because after expansion they become infrared divergent:
\bea
V_s(r) &=& 
\mu^{4-d}\int \frac{d^{d-1}k}{(2\pi)^{d-1}}e^{-i\vec k \cdot \vec r}
\left[-C_F \frac{4\pi \als }{\vec k^2}\left(1 - \frac{m_D^2}{\vec k^2} + \dots \right) 
+ i \, C_F \, \frac{T}{|\vec k|} \, m_D^2 \frac{4\pi^2\als }{\vec k^4}\left( 1 + \dots \right)
\right]
\nn\\
&=& -C_F\frac{\als}{r} - \frac{C_F}{2}\,\als\,r\, m_D^2 + \dots 
\nn\\
& & 
+ i \frac{C_F}{6} \, \als \, r^2 \, T \, m_D^2
\left(\frac{1}{\epsilon}+\gamma_E + \ln \pi + \ln (r\,\mu)^2 -1 \right) + \dots \;.
\label{VsTrmD1}
\eea
The dots stand for higher-order real and imaginary terms. In the Coulomb part,  
we have displayed only the leading term. In the imaginary part, 
the divergence comes from the Fourier transform of $1/|\vec k|^5$, which, in $d$ dimensions, 
may be found in \cite{Gelfand:1964}.
Because we expand the gluon propagator in $m_D^2/\vec k^2$ the integral corresponding to the static
quark self energy has no scale and the matching gives $\delta m=0$.

Next, we integrate out the scale $m_D$. At one-loop level this corresponds 
to evaluating the contribution to the potential of the diagram generated 
by the singlet-octet vertex (dipole interaction) in the effective theory, 
which is shown in Fig.~\ref{figEE}; the HTL resummed gluon propagator is used. 
Therefore, the contribution is the same 
as the one calculated in Sec.~\ref{secshortmD} and given in Eq.~(\ref{VsmD}).
Summing its diagonal element with Eq.~(\ref{VsTrmD1}) gives 
\bea
V_s(r) 
&=& -C_F\frac{\als}{r} - \frac{C_F}{2}\,\als\,r\, m_D^2 
+ \frac{C_F}{6} \, \als \, r^2m_D^3 + \dots 
\nn\\
& & 
-i \frac{C_F}{6} \, \als \, r^2 \, T \, m_D^2
\left(-2\gamma_E  - \ln (r m_D)^2 + \frac{8}{3} \right) + \dots \;.
\label{VsTrmD2}
\eea
We see that in the sum the divergences of Eqs.~(\ref{VsTrmD1}) and (\ref{VsmD}) cancel each 
other providing a finite physical result. The term $\displaystyle \frac{C_F}{6} \, \als \, r^2m_D^3$
in the real part is suppressed by a factor $r m_D$ with respect to $\displaystyle - \frac{C_F}{2}\,\als\,r\, m_D^2 $
and will be neglected in the following. Note the appearance of the logarithm 
$\ln (r m_D)^2$ opposed to the appearance of the logarithm $\displaystyle \ln \frac{T^2}{m_D^2}$ 
in Eq.~(\ref{sumV11}). In the first case, the logarithm signals that divergences have been canceled 
when integrating out the scales $1/r$ and $m_D$, in the second case it signals that divergences 
have been canceled when integrating out the scales $T$ and $m_D$.
The real and imaginary parts of Eq.~(\ref{VsTrmD2}) can be also obtained by expansion 
in $r m_D$ of Eq.~(\ref{energy2}) and $-\Gamma/2$, as defined in Eq.~(\ref{width2}), respectively.

\subsection{Singlet static energy for $1/r \gg m_D$}
\begin{figure}[ht]
\makebox[-9truecm]{\phantom b}
\put(0,0){\epsfxsize=7.5truecm \epsfbox{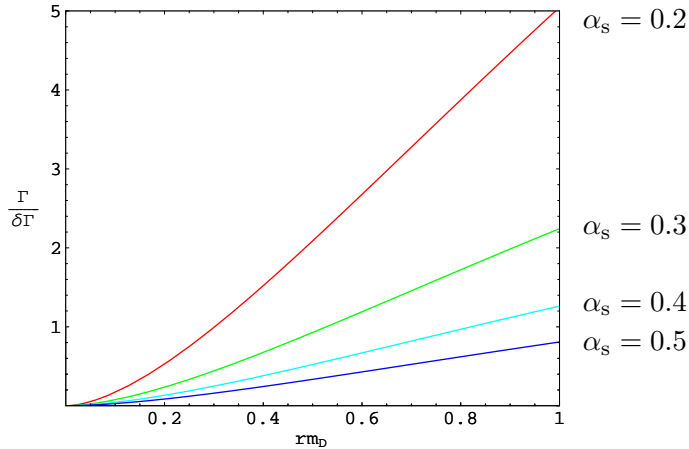}}
\put(220,160){$\als = 0.2$}
\put(220,82){$\als = 0.3$}
\put(220,53){$\als = 0.4$}
\put(220,38){$\als = 0.5$}
\caption {$\Gamma/\delta \Gamma$ vs $rm_D$ for different values of $\als$.}
\label{figRatioGamma}
\end{figure}

The real part of Eq.~(\ref{VsTrmD2}) provides the static quark-antiquark energy for $1/r \gg m_D$, whose 
leading thermal contribution is 
\be 
\delta E_s = - \frac{C_F}{2}\,\als\,r\, m_D^2 \,,
\label{energy3}
\ee
and minus twice the imaginary part of Eq.~(\ref{VsTrmD2}) provides the static quark-antiquark thermal decay width
\be 
\Gamma =  \frac{C_F}{3} \, \als \, r^2 \, T \, m_D^2
\left(-2\gamma_E  - \ln (r m_D)^2 + \frac{8}{3} \right)\,.
\label{width3}
\ee
Also in this case the singlet to octet break up process provides a contribution to the 
thermal width, which is $\displaystyle \delta \Gamma = \frac{1}{3} \, N_c^2 C_F \, \als^3\, T$.
This contribution is parametrically suppressed by a factor $(\Delta V/m_D)^2$ with respect 
to the one in (\ref{width3}). However, depending on $\als$ and $r m_D$, it may still contribute 
to a large fraction of $\Gamma$, see Fig.~\ref{figRatioGamma}.
Note that the condition $m_D \gg \Delta V$ requires $\als(1/r) \ll 2/N_c \times rm_D$ to be fulfilled. 

\section{Conclusions}
\label{secCon}
We have studied the real-time evolution of a static quark-antiquark pair 
in a medium of gluons and light quarks characterized by a temperature $T$.
We have addressed the problem of defining and deriving the potential 
between the two static sources, and of calculating their energy and 
thermal decay width. In the different ranges of temperature considered, 
we have set up and worked out a suitable sequence of effective field theories.
Our framework has been very close to the modern EFT treatment of non-relativistic and static bound states 
at zero temperature \cite{Brambilla:2004jw}, but complicated by the existence of the thermal 
scales $T$ and $m_D$. We have assumed that all the energy scales are perturbative and worked in a strict 
weak-coupling framework. This had two consequences: first, we could exploit the hierarchy
$T \gg m_D$, second, the potential that we obtained is valid in the short range.
In an EFT framework, the potential is the $r$-dependent matching coefficient that appears 
in front of the four-fermion operator that destroys and creates the bound state, after having 
integrated out all scales above the bound-state energy. Higher-order operators give lower 
energy contributions, entering into the computation of physical observables, but not 
in the Schr\"odinger equation that governs the motion of the bound state and hence are not 
of a potential type.

A systematic treatment of non-relativistic bound states in a thermal medium,
in an EFT framework and in real-time formalism, has to our knowledge not been presented so far. 
We have devoted several parts of this work to set up a proper real-time formalism for static sources.
The main outcome of this more formal aspect of the paper is in Eq.~(\ref{sumSinglet}), 
which expresses the real-time quark-antiquark propagator as an infinite sum of free propagators 
and potential or mass-shift insertions. In all the considered dynamical regimes, the structure 
of the potential is such to satisfy this equation, see Eqs.~(\ref{Vssoft}), (\ref{sumV11diag}), (\ref{diagonalm}) 
and (\ref{diagonalVs}).

We have considered a wide range of temperatures and provided 
the leading thermal effects to the potential. The results may be summarized in the 
following way. {\bf (1)}~If the temperature is smaller than or as large as the bound-state energy scale, 
there are no thermal contributions to the potential, which is simply the Coulomb potential (\ref{Vssoft}).
Thermal effects enter in physical observables in loop corrections induced by low-energy gluons. 
The static energy and the thermal width of the system have been calculated in (\ref{EsdVT11}) and (\ref{width}).
The system exhibits a thermal width, because, due to thermal fluctuations, at short range a color singlet 
quark-antiquark state may break up into an octet state and gluons. This is a novel feature 
of quark-antiquark bound states in a thermal medium that, to our knowledge, has not been considered 
quantitatively before. The singlet to octet break-up mechanism provides the dominant contribution 
to the thermal width when the temperature is as large as the bound-state energy. 
{\bf (2)}~If the temperature is larger than the bound-state energy scale but smaller than $1/r$ 
and if $m_D$ is smaller than or as large as the bound-state energy scale, then the leading thermal 
correction to the Coulomb potential is given by the sum of Eqs.~(\ref{VsT}) and (\ref{VsTloop}). 
The potential develops a real and an imaginary part. Both the singlet to octet thermal break-up mechanism 
and the imaginary part of the gluon self energy induced by the 
Landau damping phenomenon contribute to the imaginary part. If $m_D$ is smaller than the 
bound-state energy scale then the singlet to octet thermal break-up provides parametrically the dominant 
contribution to the decay width. The imaginary part of the potential originated by the imaginary part 
of the gluon self energy is divergent and needs to be regularized. 
In physical observables, the divergent part cancels against contributions coming 
from the scale $m_D$. {\bf (3)}~If also $m_D$ is larger than the bound-state energy scale, then HTL of the 
type shown in Fig.~\ref{figEE} contribute to the potential as well. The expression of the potential 
turns out to be finite at the considered order and is given by Eq.~(\ref{sumV11}). 
Equation (\ref{sumV11}) is finite because it provides the leading thermal correction 
to the static energy and to the decay width in the expansions in $\Delta V/T$, $m_D/T$ and $rm_D$. 
The same result has been also derived in perturbative QCD by calculating the Fourier transform of the 
diagrams in Fig.~\ref{figgluon00} and Fig.~\ref{figuswilson0}. 
{\bf (4)}~If the temperature is larger than $1/r$ but $m_D$ is smaller than $1/r$, the static potential is 
given by Eq.~(\ref{VsTrmD2}). If $m_D$ is also smaller than or of the same order as $\Delta V$ 
then the potential is given by Eq.~(\ref{VsTrmD1}) and divergences cancel in physical observables 
against loop corrections from the scale $m_D$. 
{\bf (5)}~Finally, if $m_D$ is of the order of $1/r$ the static 
potential is given by Eq.~(\ref{Vs}): this result agrees with the earlier finding of \cite{Laine:2006ns}.
Equations (\ref{VsTrmD2}) and (\ref{Vs}) are finite because, in the kinematical regions of validity, 
they provide the leading thermal correction to the static energy
and the decay width (see Eqs.~(\ref{energy3}),(\ref{width3}), (\ref{energy2}) and (\ref{width2})). 
In the temperature ranges {\bf (3)}, {\bf (4)} and {\bf (5)}, the parametrically dominant contribution to the 
thermal width comes from the imaginary part of the gluon self energy.

While this work was being carried out and completed some papers appeared dealing with 
some of the issues addressed here. In \cite{Kharzeev:2007ej,Lee:2008xp}, among others, the role of the 
gluon condensate was considered. The gluon condensate enters the expression of 
the mass of the quarkonium if the typical binding energy is much larger than the 
hadronic scale $\lQ$. The potential is therefore not affected, while the effect 
of the gluon condensate on the mass is parametrically smaller than the leading relativistic 
corrections. If the temperature scale is larger than the binding energy then the temperature 
enters the potential and the effects have been described in the present work.
If $T \ll 1/r$, temperature effects are carried by the dipole-dipole interaction shown 
in Fig.~\ref{figEE} and corrections to it. Only when the time scale of the chromoelectric 
correlator is smaller than the binding-energy scale, the chromoelectric correlator 
reduces to the local condensate (see Sec.~\ref{secEE}). Low temperatures, smaller than the binding 
energy, affect the condensate but not the potential, which remains Coulombic. In this case, 
thermal effects (in the static limit, they have been evaluated in Eq.~(\ref{EsdVT11condensate}))
are parametrically smaller than the leading relativistic corrections.
It remains to be clarified if a temperature scale below the binding energy is above the critical 
temperature: this may depend on the fact that the temperature scale is $T$ or a multiple of $\pi T$.
In \cite{Beraudo:2007ky}, some of the issues discussed here about the potential in  
real-time formalism have been addressed. In particular, static particles in real-time formalism 
have been considered and Eq.~(\ref{Vs}) has been derived for a hot QED plasma in a real-time 
framework. It has been also pointed out that the real part of the static potential (\ref{Vs}) 
agrees with the singlet free energy and that for large separations $r\gg 1/m_D$, $\Gamma/2$, 
as defined in Eq.~(\ref{width2}), gives twice the damping rate of an infinitely heavy quark.
Finally, in \cite{Escobedo:2008} a study of non-relativistic bound states in a hot QED 
plasma has been carried out in a non-relativistic EFT framework, which is similar to the one presented here. 

There are many possible developments of this work some of which 
have already been mentioned in the previous pages. Here we stress few of them.
First, the construction of a full EFT for non-relativistic bound states at finite temperature
requires to be completed. We have focused in the paper on the quark-antiquark 
color singlet state, but a complete identification and study of all relevant degrees 
of freedom that appear once the thermal energy scales have been integrated out, 
is still to be done. This may require the construction of an EFT that includes  
the dynamics of gauge fields below the scale $m_D$ \cite{Bodeker:1998hm}.
Second, in the EFT framework presented here, the study of quark-antiquark states 
at large but finite mass, i.e. actual quarkonium in a thermal medium, 
should be addressed. As argued along the paper, the static limit provides the first piece of a  
$1/m$ expansion; higher-order corrections may be systematically implemented in the framework of 
NRQCD and pNRQCD. Finally, although the short-distance analysis performed in this work 
may provide a valuable tool for studying the thermal dissociation of the lowest 
quarkonium resonances, the inclusion  in the analysis of the non-perturbative scale $\lQ$ 
may become necessary for studying excited states.

\begin{acknowledgments}
N.B. and A.V. would like to thank the Instituto de F\'\i$ $sica Corpuscular (IFIC) of Valencia 
for the warm hospitality and the stimulating atmosphere provided when a large part 
of this work was been carried out. N.B. and A.V. thank Miguel Escobedo Espinosa 
and Joan Soto for discussions, for reading the manuscript and for sharing their results 
with them before publication. A.V. acknowledge financial support by the IFIC. N.B.
and A.V. acknowledge financial support from the network Flavianet (EU) MRTN-CT-2006-035482.
The work of P.P. has been supported by U.S. Department of Energy under Contract No. DE-AC02-98CH10886.
\end{acknowledgments}

\end{document}